\documentclass[twocolumn,showpacs,preprintnumbers,amsmath,amssymb]{revtex4-1}

\usepackage{graphicx}
\usepackage{dcolumn}
\usepackage{amsfonts,amsmath,amssymb,bm}
\usepackage[colorlinks]{hyperref}
\usepackage{etoolbox} 

\makeatletter
\appto{\appendix}{%
  \@ifstar{\def\theequation@prefix{A.}}%
          {}%
}
\makeatother

\begin{document}

\title{\bf {Calculated Structural and Electronic Interactions of the nano
dye molecule $\rm Ru(4,\acute{4}-COOH-2,\acute{2}-bpy)_2(NCS)_2$($\rm N3$) with a iodide/triiodide redox shuttle}}
\date{\today} 
\author{E.~Shomali}
\author{I.~Abdolhosseini~Sarsari}
\author{S.~Javad~Hashemifar}
\author{M.~Alaei}
\affiliation{Department of Physics, Isfahan University of Technology, Isfahan, 84156-83111, Iran\\
}

\begin{abstract}

In this paper, dye sensitized solar cell based on nano dye molecule N3 are investigated
by using density functional computations.
The main focus is on the N3 dye molecule and corresponding complexes 
formed at the interface between electrolyte and dye, during the regeneration process.
The optimizaed geometry and electronic structure of the molecule and complexs
are calculated by using the pseudopotential as well as full-potential techniques.
The absorption spectra of metalliferous dye molecule, N3, and its complexes are
computed in the framework of time dependent density functional theory.
We determine the reaction path of dye regeneration by Nudged Elastic Band (NEB) method.
IR spectrum of the N3 dye molecule were also calculated. 
We found that complexes of N3 dye molecule and transition states formed 
in reactions, are magnetic.

\end{abstract}

\maketitle
\section{INTRODUCTION}

The world energy demand is increasing while the main energy resource, 
fossil fuels, are limited. Moreover, these conventional energy resources
are usually accompanied with various kinds of environmental pollutions. 
Therefore, renewable and clean energy sources, including wind, geothermal, 
marine and tides, biomass, and solar energy \cite{panwar2011role}
are strongly desired. 
Solar photovoltaic is likely the leading renewable energy technology 
which is expected to provide an energy share of about 16\% in 2050
based on Technology Roadmap Solar Photovoltaic Energy, 2014.
First generation solar cells are based on crystalline silicon that 
are stable and have a high efficiency but their construction is costly. 
Second generation solar cells, which employ non-crystalline silicon,
have lower efficiency and cost than the first generation. 
Third generation solar cells are based on organic materials, 
consisting of dye sensitized solar cells, polymer solar cells, and 
liquid-crystal solar cells \cite{Bagher2015,Green2014,crabtree2007solar}. 
Although dye sensitized solar cells have low efficiency (almost 11.1 \%), 
their low cost and acceptable lifetime (at least 20 years under operational conductions) 
have attracted great attention \cite{gratzel2006photovoltaic}. 
These solar cells are made of photoanode that is a mesoporous oxide film combined with 
semiconductor particles at the nanometer-scale. The semiconductor particles that are protected 
by a transparent conductive oxide are usually TiO$_2$. Dye molecule that is placed on the 
photoanode, absorbs photon flux emitted from the sun. Also, it can protect the contacting of 
TiO$_2$ surface with the electrolyte solution and consequently prevents from charge 
recombination process \cite{pastore2012computational}. Afterward the photo-excited electron from 
dye molecule is enjected into the conduction band of semiconductor and through the distribution 
of electron in it, current is generated. The electron proceeds to the counter electrod that is 
placed in adjacency of a electrolyte solution. Then redox couple in the solution transfers 
electron to dye molecule and eventually dye molecule is regenerated to ground state 
\cite{Bagher2015,Gratzel2003}.

\begin{figure}[!ht]
\centering
\includegraphics*[scale=0.2]{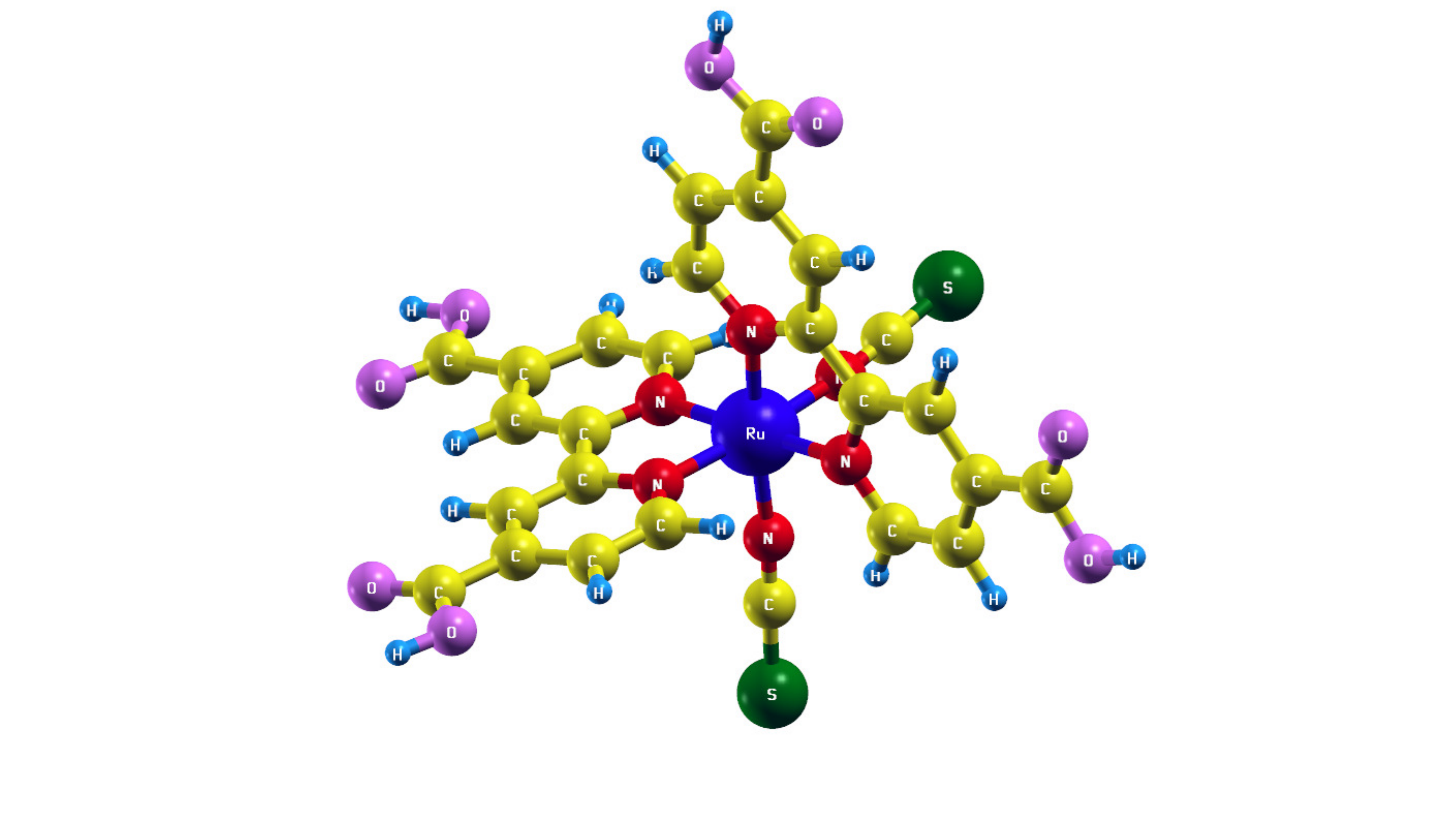}
\caption{ 
  Optimized structure of N3 dye. The blue, red, yellow, violet, 
  green and light blue spheres represent Ru, N, C, O, S and H atoms, 
  respectively.
}
\label{N3}
\end{figure}

In this paper, dye sensitized solar cells based on  
N3 dye molecule (which has been the subject of a combined experimental and theoretical study 
\cite{pizzoli2012acid}), with an electrolyte of iodide/triiodide redox 
couple are investigated. 
The N3 dye molecule is placed in category of Ruthenium Polypyridyl sensitizer dyes
and has 59 atoms including Ru, N, S, C, O and H atoms \cite{Gratzel2003}. 
The optimized molecular structure of N3 is displayed in Fig. \ref{N3}. 
This molecule absorbs incident sunlight photon in a range of visible spectrum, 
has a high thermal stability and excited states with high lifetime which have led 
to be recognized as an efficient sensitizer. 
In addition, the acid carboxylic group (COOH) used to putting dye on the surface of 
the semiconductor \cite{de2004time}. Redox couple in the electrolyte solution, 
iodide/triiodide, has an acceptable redox potential, is soluble and accelerates 
the dye regeneration process. 
On the other hand, its slow exchange current density on the surface of TiO$_2$ 
leads to reducing unwanted charge recombination on the interface of TiO$_2$ and 
redox couple (\textquoteleft the dark current\textquoteright). 
Therefore, due to these characteristics, $\rm I^-/I_3^-$ redox couple 
is superior than others \cite{boschloo2009characteristics}.

In some studies by Clifford et al., dye regeneration process has been suggested through 
the formation of an unstable intermediate complex according to the following reactions:

\begin{equation}
 \rm Dye^+ + I^- \longrightarrow [Dye^+ ... I^-]
\end{equation}
\begin{equation}
 \rm [Dye^+ ... I^-] + I^- \longrightarrow Dye + I_2^-
\end{equation}

This process is accompanied by liberalization of diiodide ($\rm I_2^-$) 
\cite{Asaduzzaman2011,clifford2007dye}.
The unstable intermediate complexes [$\rm Dye^+ ... I^-$] can be formed in several ways,
that depends on the bonding of iodine atom with dye molecule. 
Here, we have examined two more likely [$\rm N3^+ ... I^-$] complexes. 
In the first one, [$\rm N3^+ ... I_{Ru}^-$], the iodine atom 
is located between two pridine (dcbpy) rings, 
while bonding of iodine to the hydrogen atom of carboxylic group (COOH)
give rises to the second more likely complex, [$\rm N3^+ ... I_{OH}^-$].
For as much as [$\rm N3^+ ... I^-$] 
intermedium is still able to absorption of another iodide ion, $\rm N3I_{Ru}-I^-$ and $\rm N3I_{OH}-I^-$ 
complexes construct, that in them the second $\rm I^-$ is interacted with first $\rm I^-$ of 
[$\rm N3^+ ... I_{Ru}^-$] and [$\rm N3^+ ... I_{OH}^-$] respectively \cite{Asaduzzaman2011}, 
\cite{xie2014theoretical}.
First-principles investigation of the structural, chemical, and energetic properties 
of these complexes would be helpful for better perception of the dye regeneration process. 
Moreover, we will also determine the minimum energy path for the proposed reactions.

\section{Computational Details}

Our electronic structure calculations and geometry optimizations 
have been performed in the framework of density functional theory (DFT) 
by using PAW pseudpopotential as well as numerical orbital - full potential 
techniques implemented in the Quantum Espresso (QE) \cite{Giannozzi2009} and FHI-aims \cite{Blum2009}
computational packages, respectively.
We used GGA-PBE exchange-correlation functional, \cite{PhysRevLett.77.3865}
and supercell appraoch for simulation of isolated dye molecules in QE.
A vacuum thickness of about 20 Bohr was utilized to avoid interaction
of adjacent moleucles. 
Enargy cutoffs of 30 Ry and 300 Ry, were used for plane wave expansion of wave functions
and electron density, while full potential calculations were perfromed with
$tier2$ basis set and atomic ZORA scalare relativistic effects. 

The optical absorption spectrum of the molecules were calculated in 
the framework of Liouville-Lanczos approach to time-dependent DFT, implemented 
in Turbo-TDDFT code \cite{malciouglu2011turbotddft,ullrich2014brief}, 
that is part of the Quantum Espresso distribution.
The Nudged Elastic Band (NEB) method \cite{Caspersen2005}, implemented in Quantum Espresso,
was also used to determine the reaction path.

\section{RESULTS AND DISCUSIONS}

\subsection{Structural properties}

\begin{figure}[!ht]
\centering
\includegraphics*[scale=0.5]{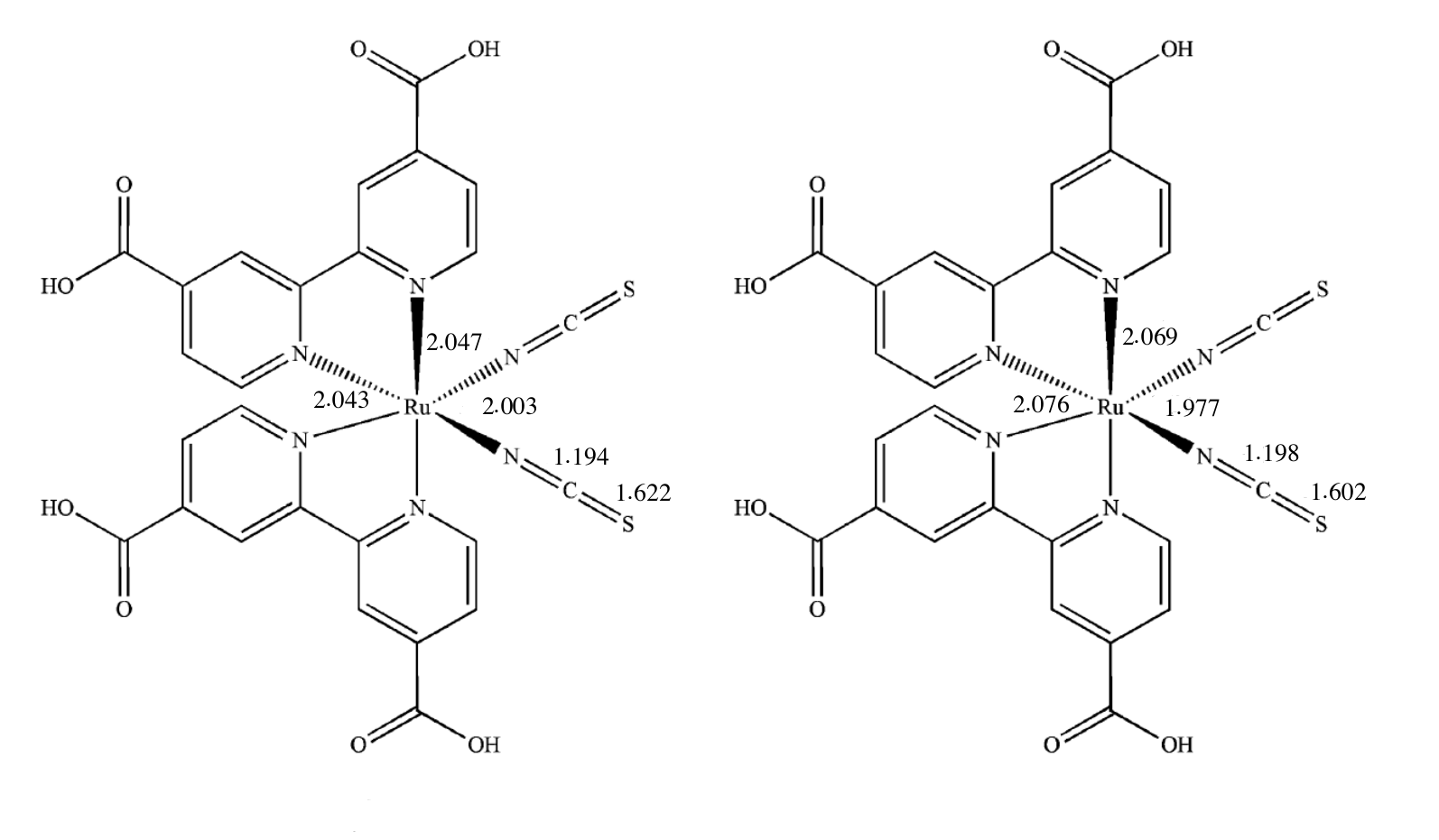}
\caption{
  Schematic representation of the optimized structure of N3 dye 
  and its oxidized form, N3$^+$.
  The numbers in the figure, show the corresponding bond length in Angstrom.
}
\label{N3-N3}
\end{figure}

As already mentioned in dye regeneration mechanism, the $\rm N3$ dye is excited by absorbing 
photons, and turns to the $\rm N3^+$. Afterwards, interaction of the $\rm N3^+$ with the iodide 
redox shuttle cause the dye regeneration. 
By repeating this cycle processes, current is generated. 
Hence, the oxidized dye ($\rm N3^+$) plays a crucial role in the mechanisms that occur 
in the electrolyte. The atomic bond legths of N3 and its oxidized form, N3$^+$, 
are shown in Fig.~\ref{N3-N3}. The calculated bond lengthes are in 
agreement with the reported findings by Asaduzzaman et al. \cite{Asaduzzaman2011}.

\begin{figure}[!ht]
\centering
\includegraphics[scale=1.3]{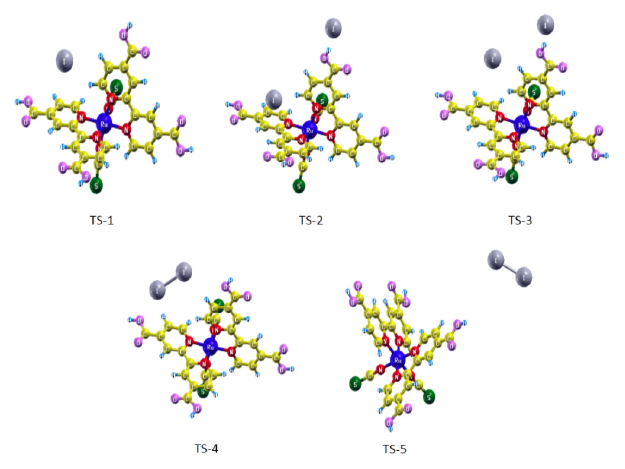}
\caption{
  Optimized structure of transition states. $\rm TS-1$ corresponds to $\rm N3^+I^-$, and other 
  transition states correspond to $\rm N3I_2^-$. Also $\rm TS-2$ and $\rm TS-4$ are related to 
  [$\rm N3^+ ... I_{Ru}^-$] while $\rm TS-3$ and $\rm TS-5$ are related to [$\rm N3^+ ... I_{OH}^-$].
}
\label{TS}
\end{figure}

The values indicate that oxidation of N3 increases the bond length between Ru and 
Pridine (dcbpy) rings and decreases the bond distance between Ru and NCS groups. 
Actually, the Ru atom in N$3^+$, compared with N3, 
has one less electron for covalent bonding with dcbpy rings, 
hence the corresponding bond distance become longer. 
On the other hand, increasing the charge on Ru atom, enhances and consequently shorten
the ionic bonding between Ru and NCS groups.

As previously mentioned, $\rm I^-$ may interact with the electropositive parts of $\rm N3^+$ 
to construct [$\rm N3^+ ... I^-$] complexes. Ru atom has a positive charge in $\rm N3^+$, but it has 
already been occupied with six bonds. So $\rm I^-$ could not interact with Ru atom and for this 
reason, $\rm I^-$ interacts with the dcbpy rings to form [$\rm N3^+ ... I_{Ru}^-$] complex.
Also, there are two options for interaction of the $\rm I^-$ with H atoms of $\rm N3^+$: The 
$\rm I^-$ can interact with H atoms, those related to the dcbpy rings and those related to the 
carboxylic groups (COOH). For as much as the H atoms that are part of the carboxylic groups, 
are more electropositive, their interaction with $\rm I^-$ leads to the formation of the more 
stable forms of [$\rm N3^+ ... I_{OH}^-$] complex. In the next step, the second iodide ion 
interacts with the first iodide ion in [$\rm N3^+ ... I_{Ru}^-$] and [$\rm N3^+ ... I_{OH}^-$] 
complexes, so $\rm N3I_{Ru}-I^-$ and $\rm N3I_{OH}-I^-$ complexes are constructed, respectively. Also, 
it is possible to the second iodide ion interacts with the H atoms of different carboxylic 
groups (COOH), but the complexes formed in this way are less stable \cite{Asaduzzaman2011}. 
Therefore, we have not considered these unstable complexes for this article. 
The stable complexes are formed through the reactions would mention in appendix
that depending on the status of iodine in complexes have been examined separately.

Theoretical efforts to achieve the transition states of the formation 
of [$\rm N3^+ ... I_{Ru}^-$] and [$\rm N3^+ ... I_{OH}^-$] have failed \cite{Asaduzzaman2011}.
These efforts only led to finding a transition state that is created during the 
conversion of [$\rm N3^+ ... I_{Ru}^-$] to [$\rm N3^+ ... I_{OH}^-$]. In other words, iodine has moved 
from its place in [$\rm N3^+ ... I_{Ru}^-$] to that in [$\rm N3^+ ... I_{OH}^-$] through the transition 
state that was named $TS-1$. The next transition states are obtained via the mentioned 
reactions \cite{Asaduzzaman2011}. The optimized structures of transition states are shown in 
Fig. \ref{TS}. The $\rm Ru-I$, $\rm H-I$ and $\rm I-I$ distances of the $\rm N3^+I^-$ and $\rm N3I_2^-$ 
complexes and mentioned transition states are tabulated in table~\ref{data}.

\begin{table*}[!ht]
\caption{
  Calculated bond lengthes (\AA), total magnetization $\mu$ ($\mu_B$),
  and optical gap of the N3 molecule and its derivatives.
  The presented bond lengthes in the paranthesis are taken from ref. \cite{Asaduzzaman2011}.
}
\label{data}
\begin{ruledtabular}
\begin{tabular}{cccccc}
                      &      \multicolumn{3}{c}{Bond length}       &$\mu$&  gap \\
\cline{2-4}
                      &     Ru - I   &     H - I    &      I - I   &     &      \\
\hline
$\rm N3$                  & ---          & ---          & ---          & 0.0 & 1.56 \\
$\rm [N3^+ ... I_{Ru}^-]$ & 5.237 (5.234)& ---          & ---          & 1.0 & 1.67 \\
$\rm [N3^+ ... I_{OH}^-]$ & ---          & 2.370 (2.392)& ---          & 1.0 & 1.65 \\
$\rm N3I_{Ru}-I^-$        & 5.639 (5.637)& ---          & 3.193 (3.265)& 1.0 & 1.50 \\
$\rm N3I_{OH}-I^-$        & ---          & 2.394 (2.462)& 8.954 (3.062)& 1.0 & 2.24 \\
$TS-1$                & 6.193 (8.500)& 6.404 (4.043)& ---          & 1.0 & 1.62 \\
$TS-2$                & 5.429 (6.031)& ---          & 8.299 (5.000)& 1.0 & 1.73 \\
$TS-3$                & ---          & 2.350 (2.432)& 5.838 (5.500)& 1.0 & 1.65 \\
$TS-4$                & 8.183 (8.200)& ---          & 2.940 (3.015)& 1.0 & 1.78 \\
$TS-5$                & ---          & 5.992 (6.000)& 17.25 (2.935)& 1.0 & 1.66 \\
\end{tabular}                                         
\end{ruledtabular}
\end{table*}

After optimization, the magnetic properties of the system were investigated. 
For this purpose, using optimized structures and considering an initial moment on iodine,
we performed spin-polarized calculations to determine the total magnetization 
of N3 and its derivatives, listed in table~\ref{data}.
It is seen that the N3 dye molecule is non-magnetic while 
its studied derivatives have a spin moment of 1 $\mu_B$. 

\subsection{Electronic and Vibrational properties}

The electronic structure of the optimized N3 molecule was caculated 
by using both pseudopotential and full potential techniques and 
a good agreement was observed.
The HOMO-LUMO gap of N3 was found to be 0.64 eV (0.61 eV) within pseudo-poetntial 
(full-potential) method, which is significantly lower than 
the experimental gap of about 1.67 eV \cite{Bersch2008}.
This difference reflects the main deficiency of LDA/GGA functionals
for prediction of the excited states properties.
Thereby, in order to achieve reliable results for energy gap, 
TDDFT calculations will be used.

\begin{table*}[htb]
\caption{\label{HL} 
Orbital density of N3 dye molecule and its derivatives; $\rm [N3^+...I_{Ru}^-]$, 
$\rm [N3^+...I_{OH}^-]$ and $\rm [N3I_{Ru}-I]$; in HOMO and LUMO states.
}
\begin{ruledtabular}
\begin{tabular}{ccc} 
Phase      &     a)  $\rm N3$                               &   b) $\rm [N3^+...I_{Ru}^-]$   \\
\hline
LUMO       &  \includegraphics[scale=0.06]{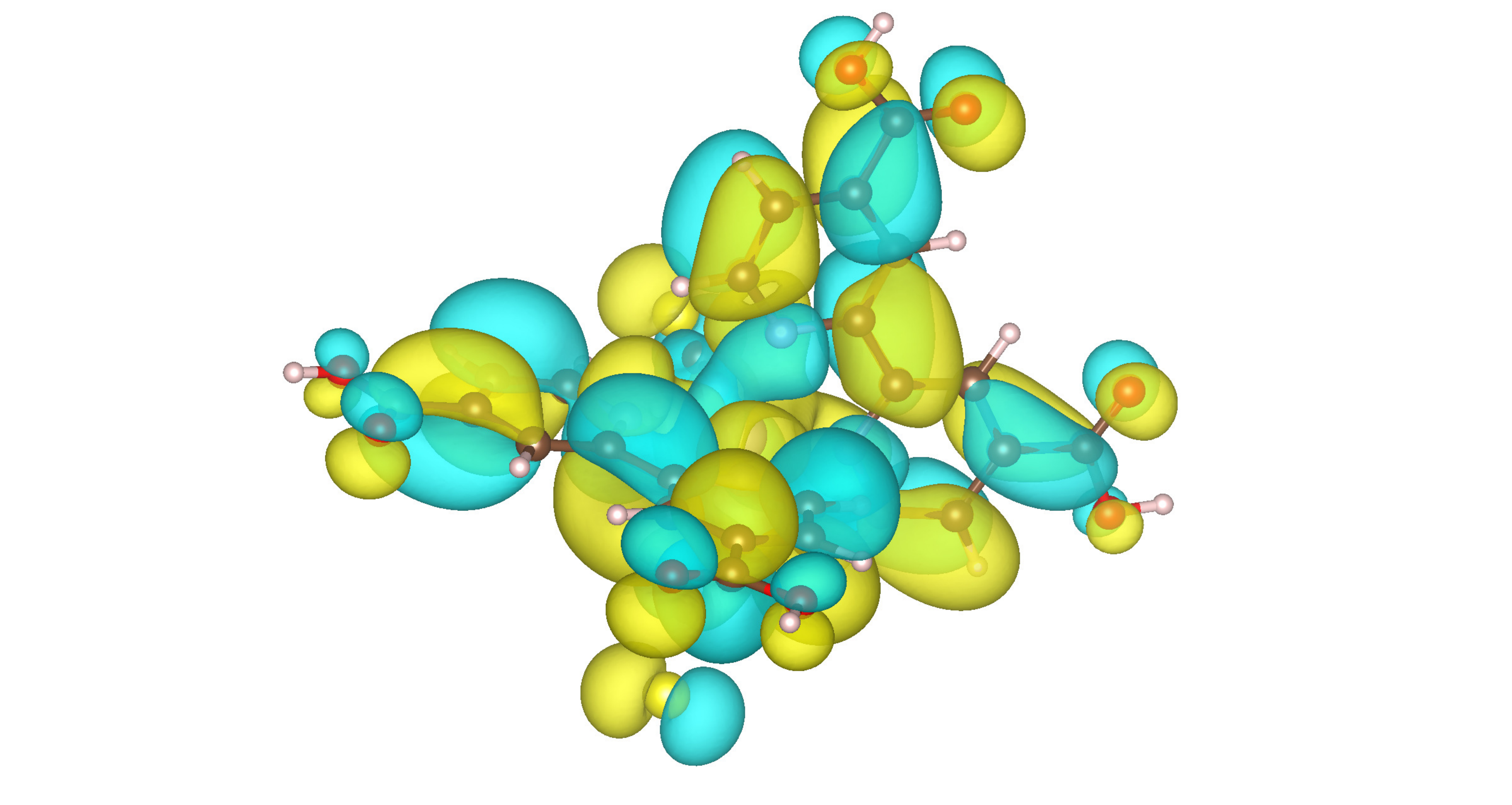} &  \includegraphics[scale=0.06]{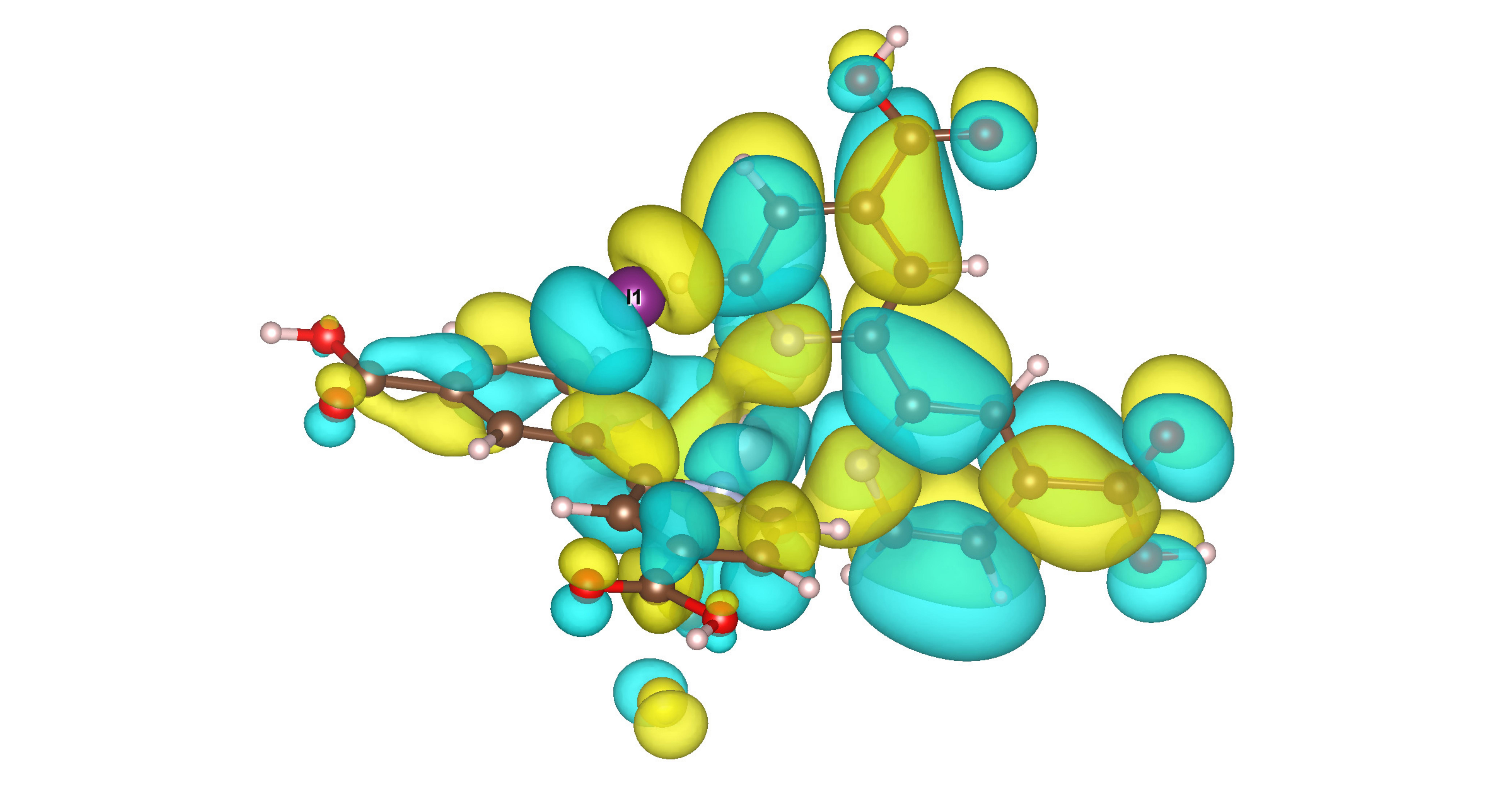}   \\
HUMO       &  \includegraphics[scale=0.06]{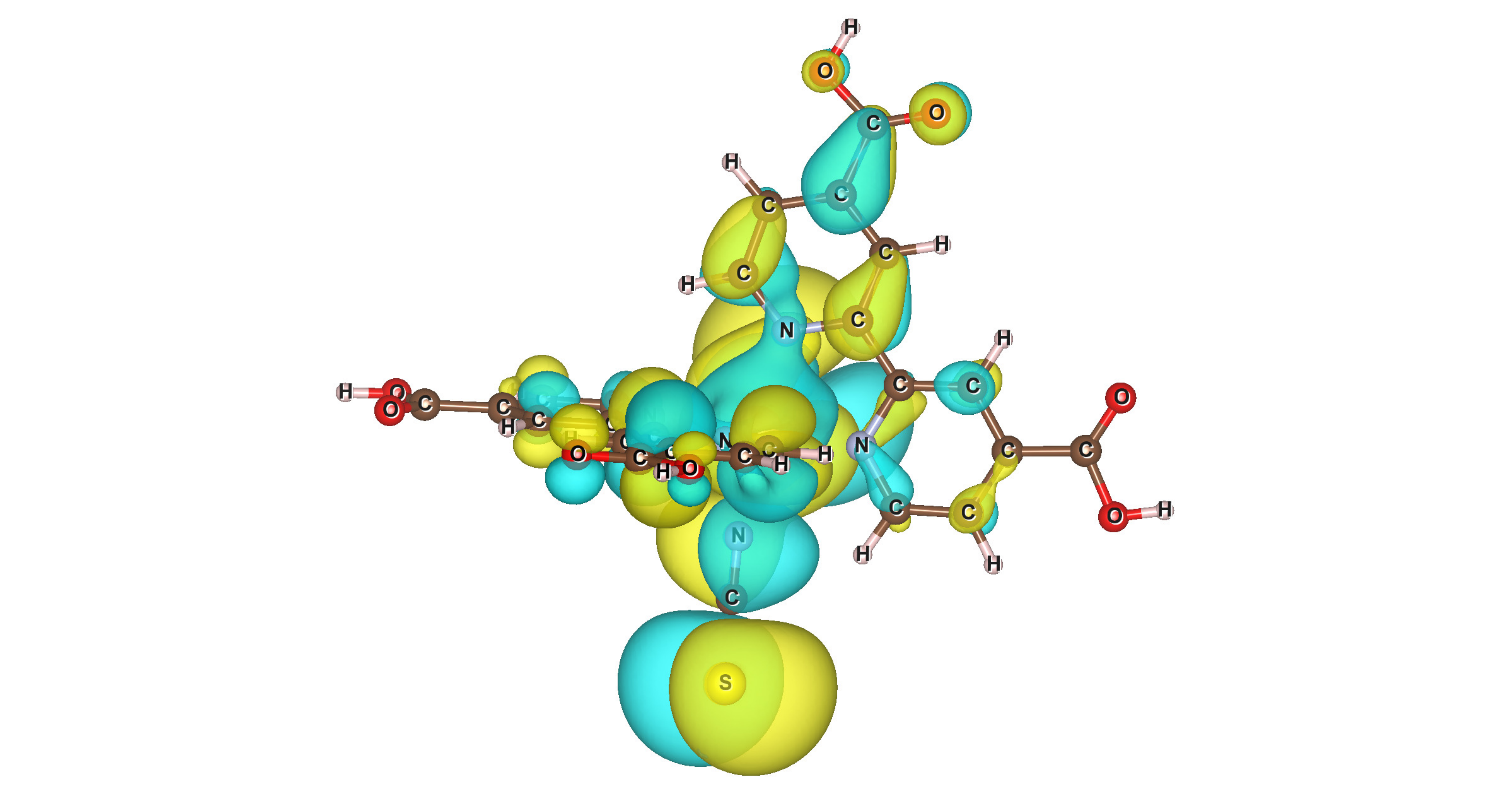} &  \includegraphics[scale=0.06]{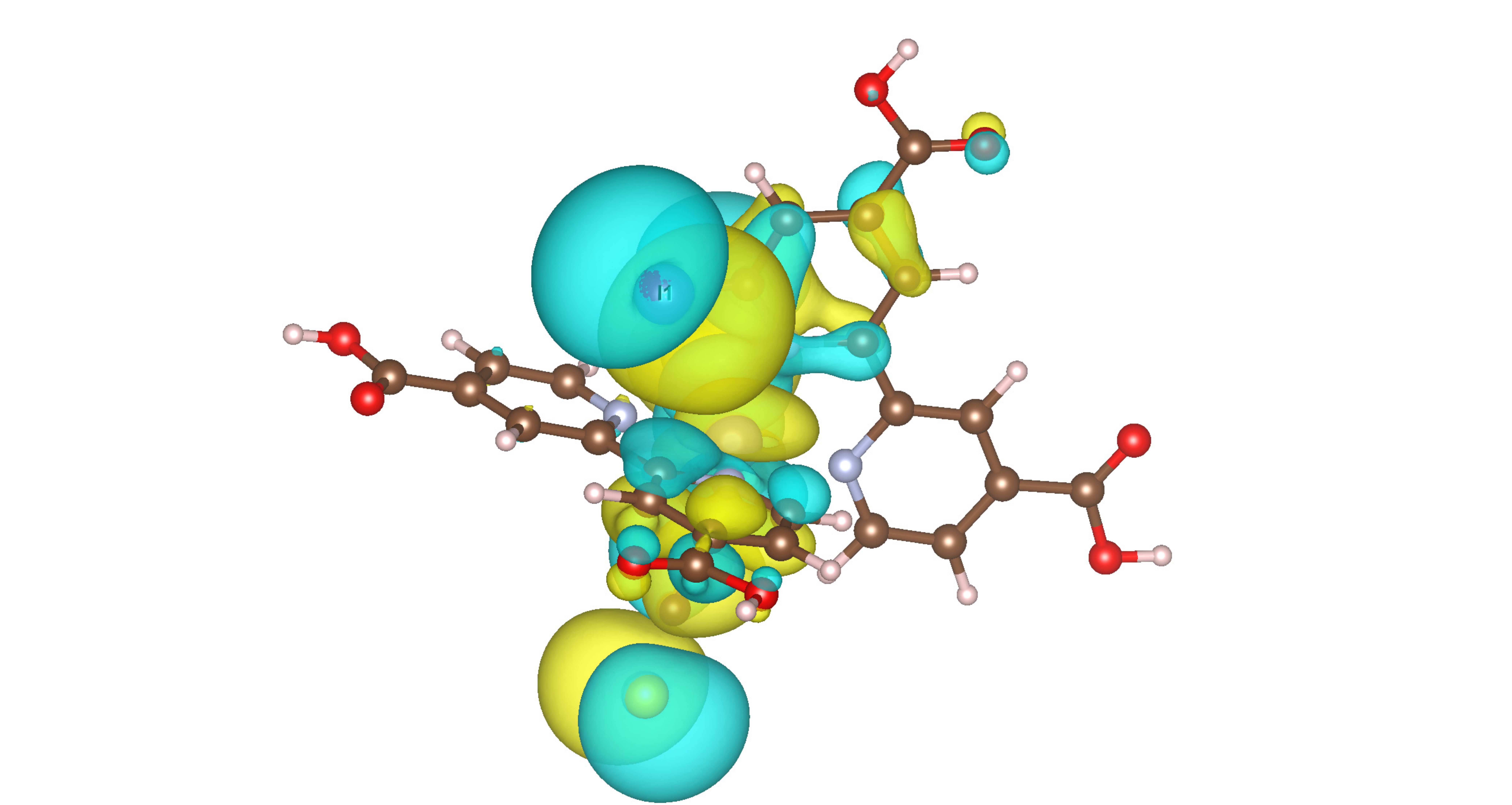}   \\
\hline
\hline
Phase      &     c) $\rm [N3^+...I_{OH}^-]$        &   d)  $\rm [N3I_{Ru}-I]$                \\
\hline
LUMO       &  \includegraphics[scale=0.06]{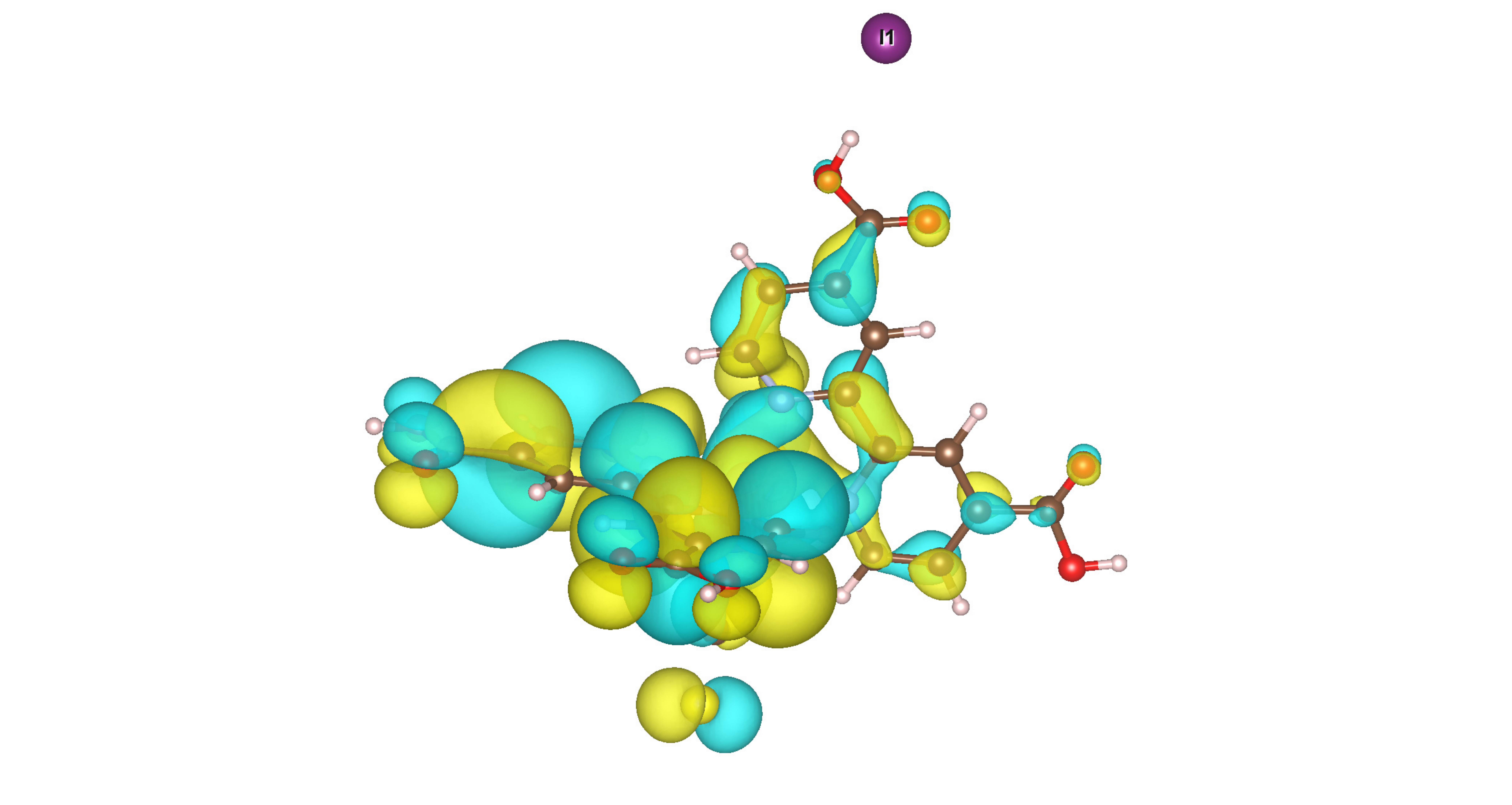} &  \includegraphics[scale=0.06]{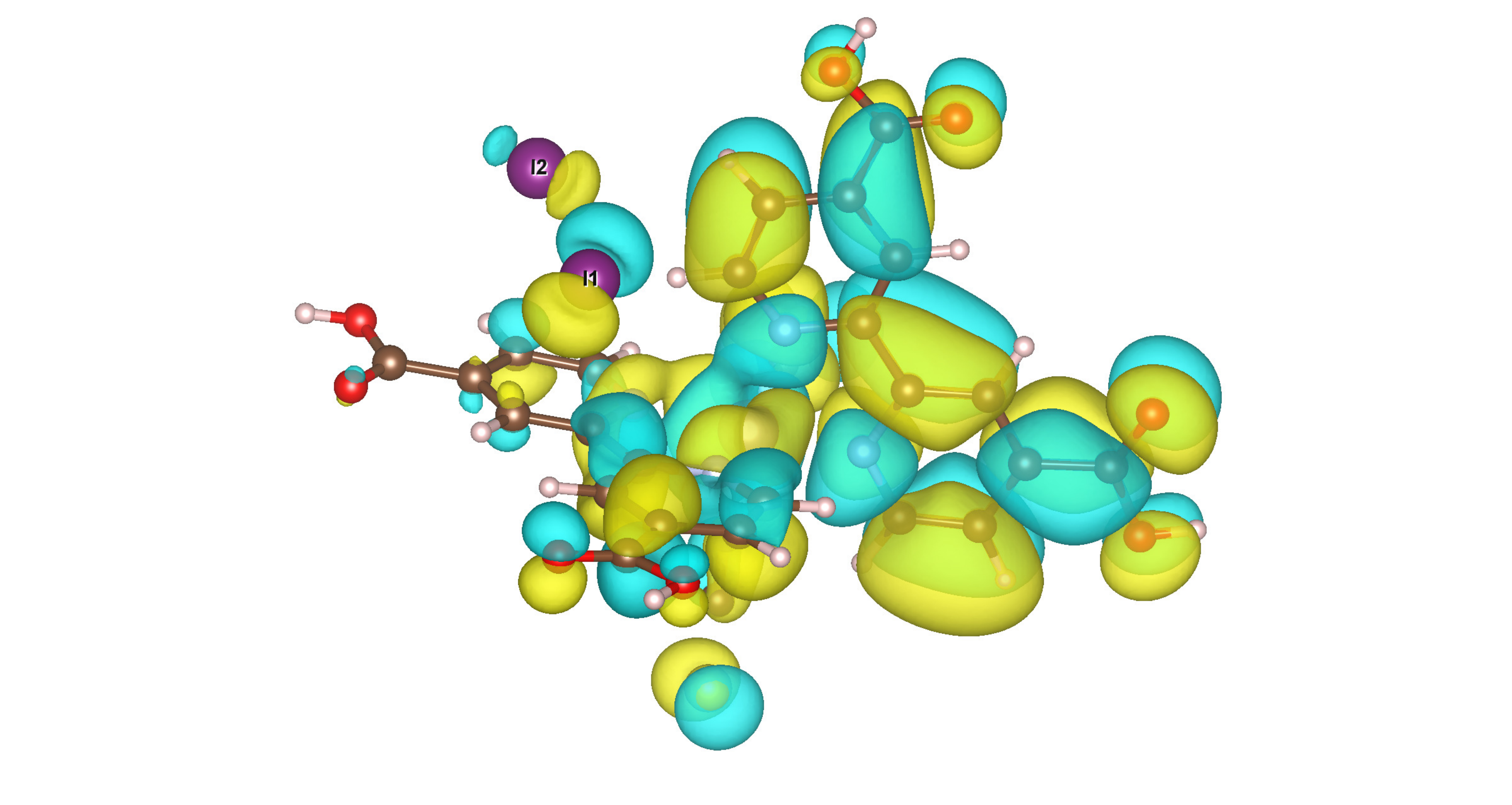}   \\
HUMO       &  \includegraphics[scale=0.06]{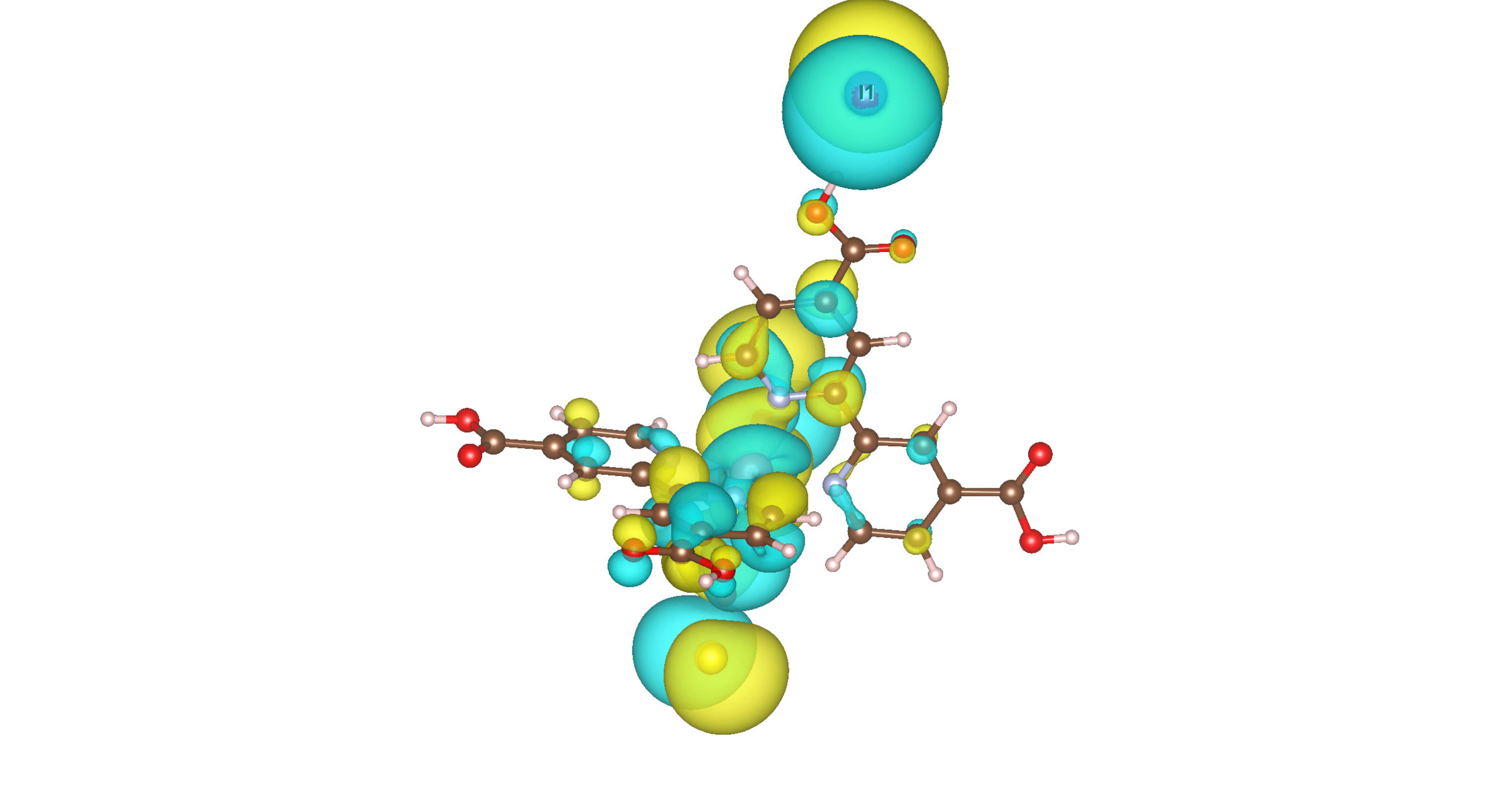} &  \includegraphics[scale=0.06]{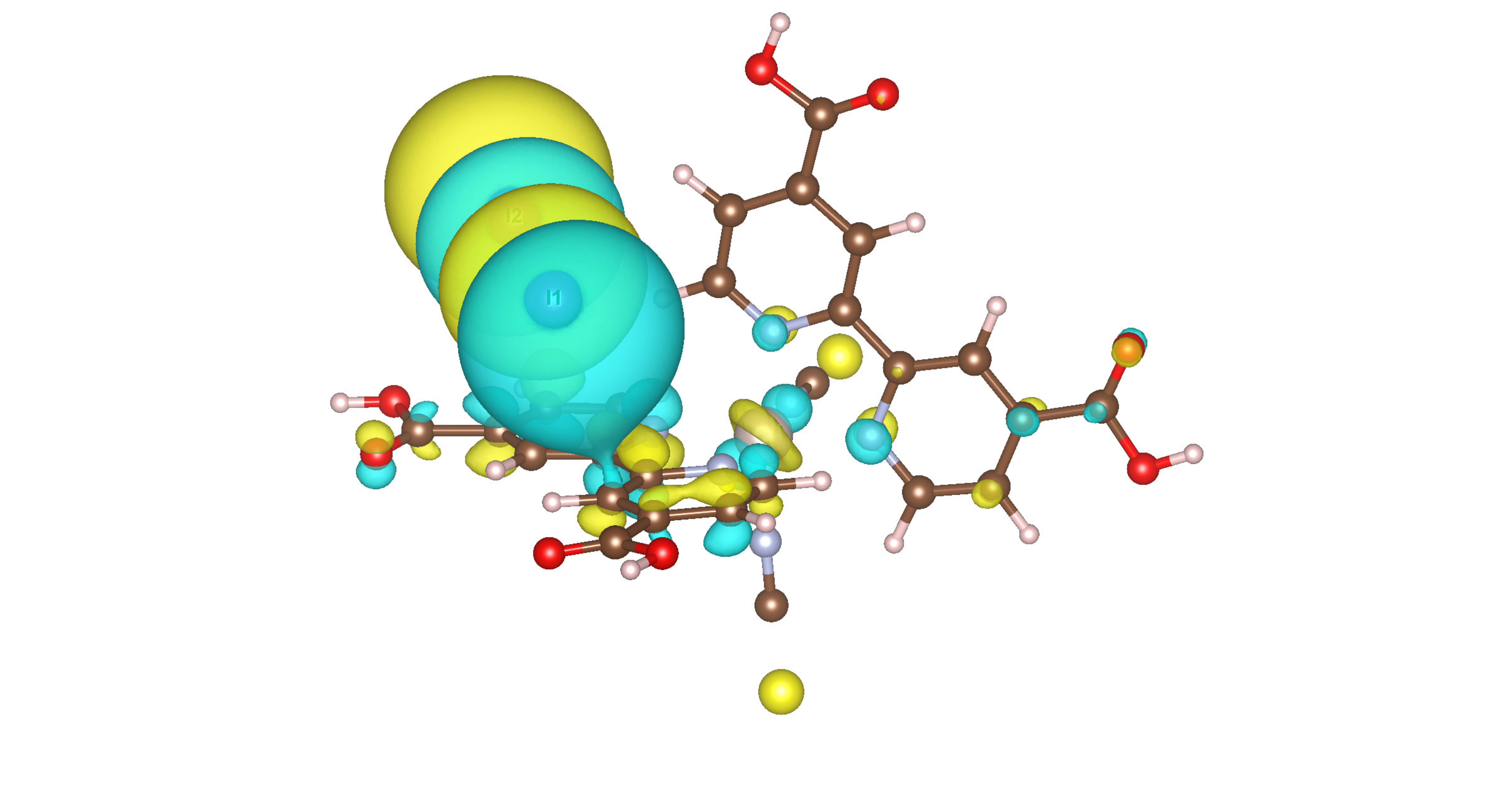}   \\
\end{tabular}
\end{ruledtabular}
\end{table*}

\begin{figure}[!ht]
\centering
\includegraphics*[scale=0.55]{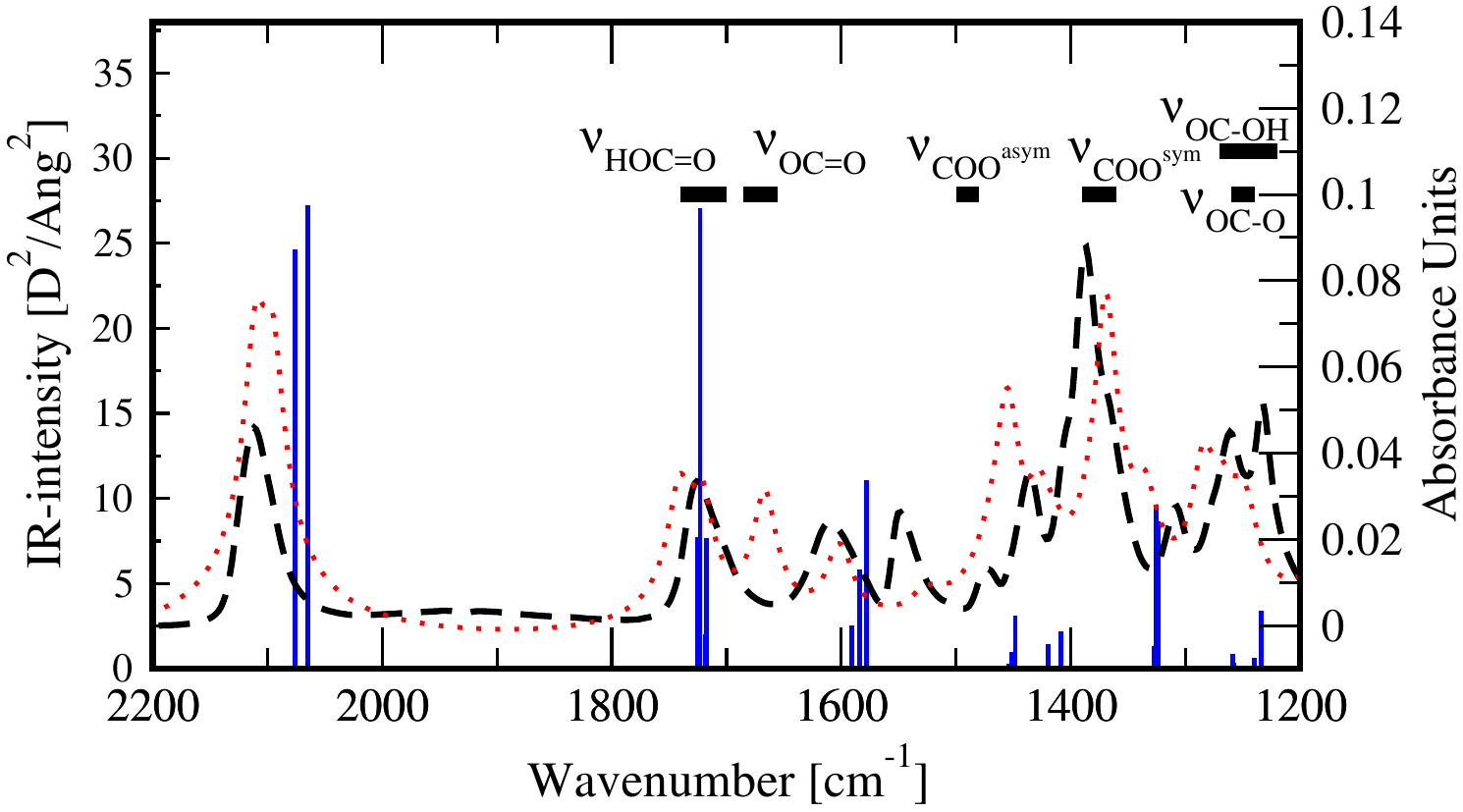}
\caption{\label{ir}
 Blue line: calculated IR spectrum of $\rm N3$ dye molecule (this work).
experimental IR spectrum of N3 adsorbed on anatase nanocrystals 
         in EtOH (black dashed) and the computed IR spectrum (red dotted) for the fully protonated 
         configuration $\rm I_1$ on the anatase TiO$_2$ (101) surface \cite{Bai2014,Schiffmann2010}).
}
\end{figure}

Orbital density of N3 molecule and some of its derivatives were calculated and 
displayed in table \ref{HL}. The most activity of electrons are in HOMO state. 
So we are enthusiastic to investigate the HOMO-LUMO orbitals of each structure. As shown 
in table \ref{HL}, the orbital density is more in HOMO orbitals than others and there are 
a lot of active electrons here, that are ready to charge exchange and reactions. This fact 
confirms the reaction of iodine atom in complexes.

The vibrational spectra of the molecule was calculated to investigate
dynamical stability and IR spectrum of the systems. 
The obtained IR spectrum of the N3 molecule is shown in Fig. \ref{ir}.
Absence of any imaginary mode in the spectrum, confirms dynamical
stability of the molecule.
As is evident, vibrations around 2100 $(cm^{-1})$ and 1700 $(cm^{-1})$ frequencies have 
higher IR intensity and they are related to the stronger bonds.
For comparison, the computational and experimental infrared spectrum of 
the N3 molecule adsorbed on TiO$_2$ anatase, taken from ref. \cite{Bai2014},
are also displayed in Fig.~\ref{ir}. 
We observe that the two highest energy peaks of our spectrum, are also visible 
in the spectrum of the adsorbed molecule, while the lower energy peaks
does not mach correctly.
It may be attributed to the presence of TiO$_2$ substrate.

\subsection{Optical properties}

\begin{figure}[!ht]
\centering
\includegraphics*[scale=0.55]{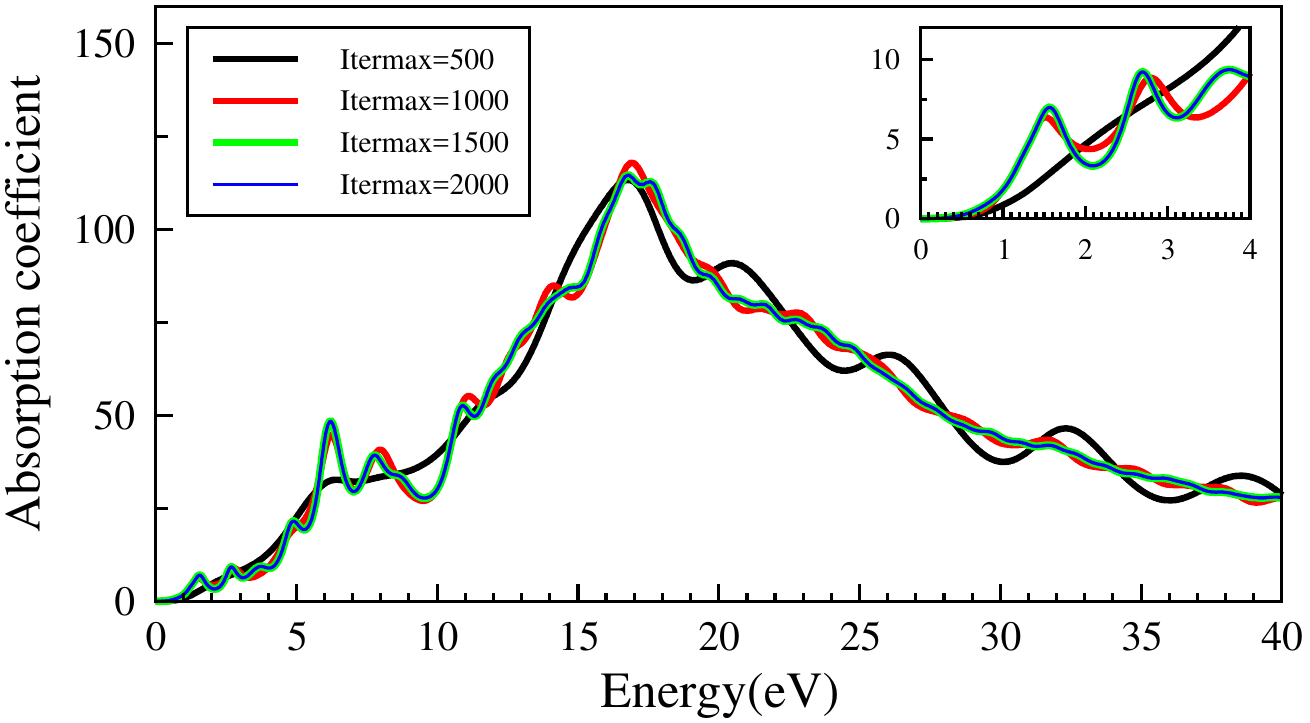}
\caption{\label{iter}
  Optimizing the number of iterations to determine 
  the optical spectrum of the N3 dye molecule.
}
\end{figure}

Absorption spectra of dye molecule play a significant role in 
the efficiency of solar cells, hence first-principles calculation 
of this property is desired.
To determine the optical spectrum, the Liovile-lanczos appraoch 
with TDFT was used. 
In this approach, an important parameter is number of iterations
which should be optimized. 
We have calculated the absorption spectrum of N3 
at different number of iterations (Fig.~\ref{iter}). 
It is seen that the optical spectrum with 500 iteration is flat 
while sharp peaks appear by increasing iterations. 
The obtained absorption spectra with 1500 and 2000 iterations are completely overlapping. 
Therefore, 1500 iterations is the optimized parameter for our subsequent calculations.

 \begin{figure}[!ht]
\centering
\includegraphics*[scale=0.55]{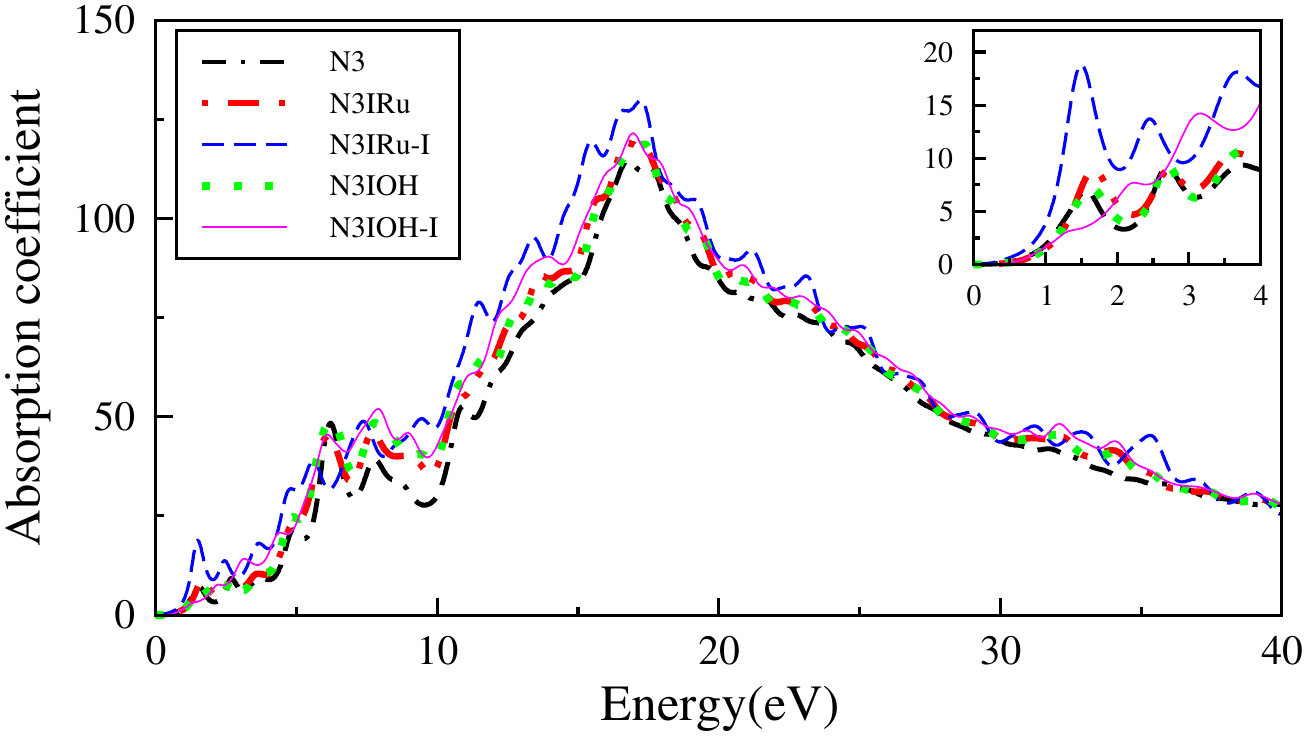}
\includegraphics*[scale=0.55]{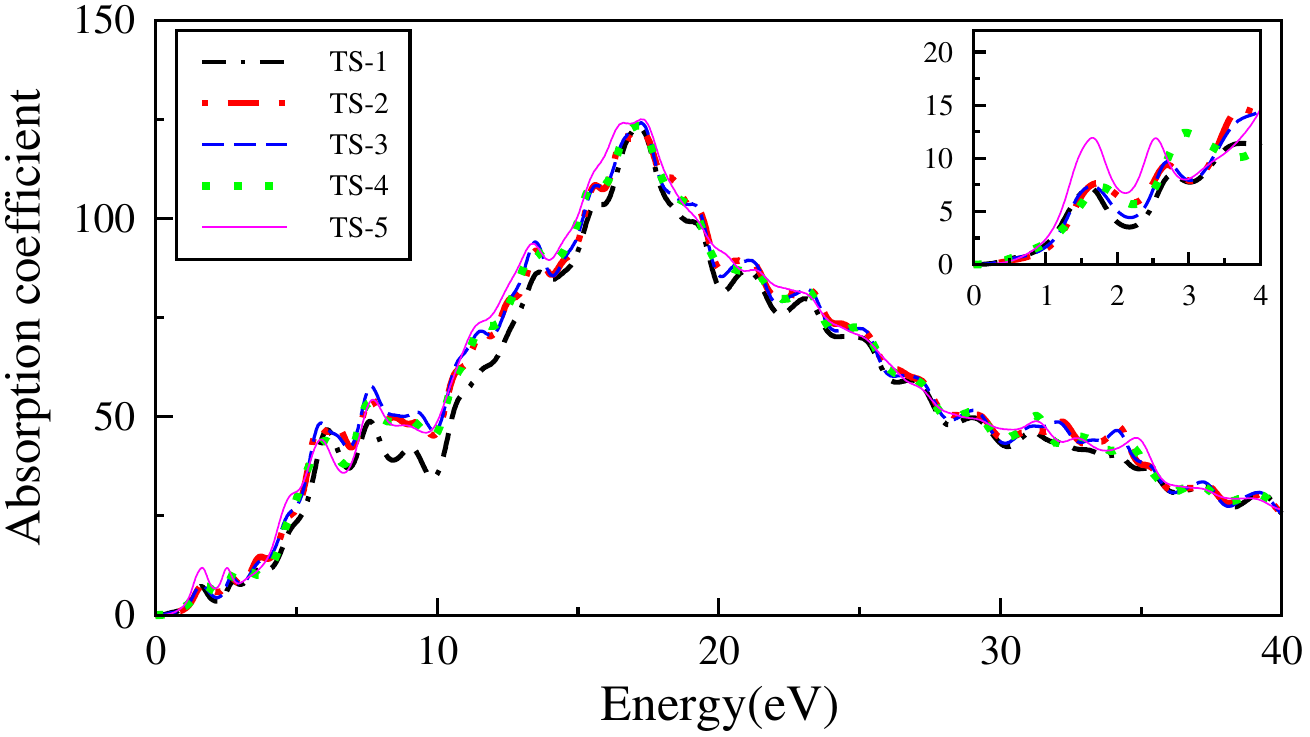}
\caption{\label{optic}
  Obtained absorption optical spectrum of N3, its complexes, 
  and corresponding transition states.
  The insets show the initial part of the spectra in a narrow window.
}
\end{figure}

\begin{figure}[!ht]
\centering
\includegraphics[scale=0.55]{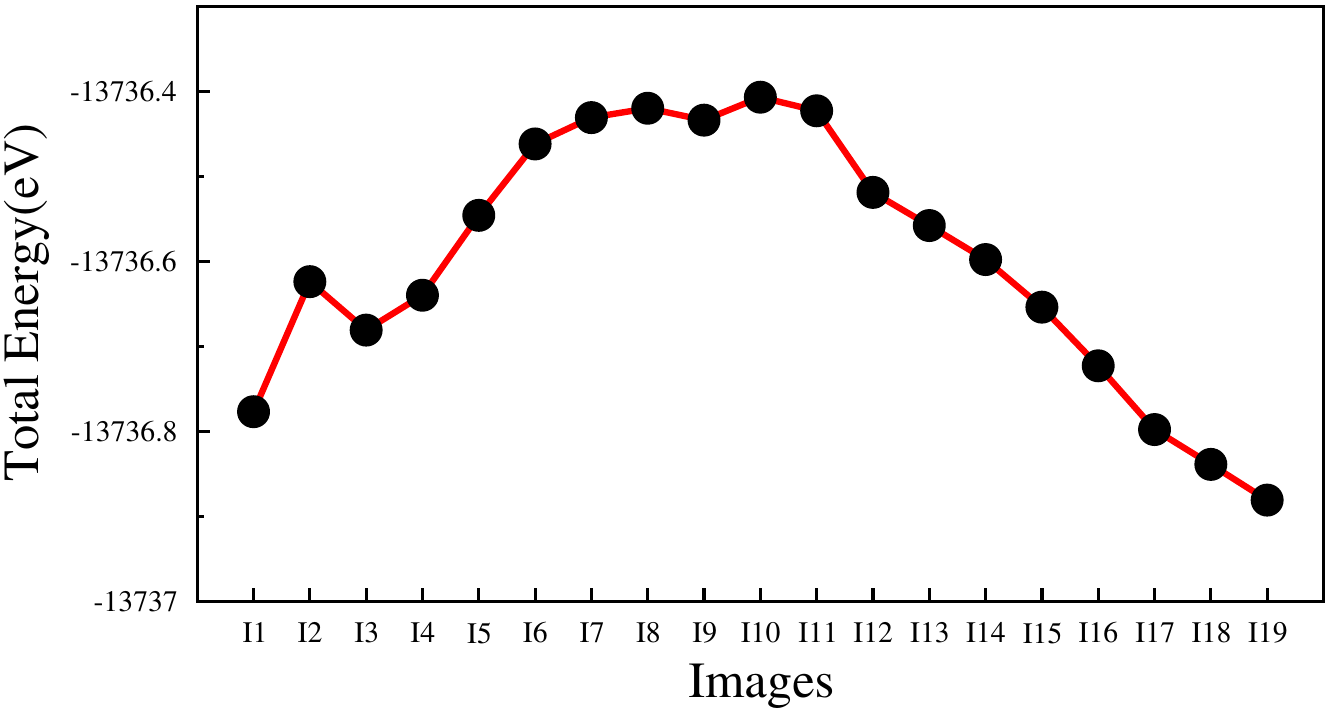}
\caption{\label{nebpath}
 Obtained minimum energy path for reaction \ref{equ:TS-1}, 
 The 10th image is considered as the transition state where I1 stands for
 $\rm [N3^+...I_{Ru}^-]$ and I19 for $\rm[N3^+...I_{OH}^-]$.
 }
\end{figure}

The optical spectrum of N3 molecule, its complexes, and the transition states
 are shown in Fig.~\ref{optic}.
The first peak in the absorption spectrum represents the first excited state and its energy is 
equal to the required energy for the transition of an electron from the highest occupied 
molecular orbital (HOMO) to the lowest unoccupied molecular orbital (LUMO), which is indeed 
representative of the optical gap. The next peaks in the absorption spectrum show the next 
excitation of electrons from other states. By increasing the energy difference between states, excitation 
occurs in higher energy. The first peak is only related to one excitation (first excitation), 
while the next peaks are related to some excitations which broadening of each peak depends on 
the number of electron excitations in the corresponding energy. The optical gap is calculated 
for structures and the results are listed in table~\ref{data}. The optical gap of $\rm N3$ 
molecule is compared with its experimental gap and the conformity of them is perfectly clear.

\begin{table}[!ht]
\begin{center}
\caption{
The required activation energy for the reactants in which transition states is formed.
}
\label{tab:activationE}
\begin{tabular}{ccccccccl}
\hline\hline
structure                   & $E_a (NEB)$ & $E_a (Quantum-Espresso)$ & $E_a$ (\cite{Asaduzzaman2011}) \\
\hline
$TS-1$                      & 8.52        & 5.68            & 7.33     \\
$TS-2$                      & -           & -               & 4.18     \\
$TS-3$                      & -           & -               & 9.57     \\
$TS-4$                      & -           & 4.18            & 5.33     \\
$TS-5$                      & -           & 3.44            & 5.20     \\
\hline
\end{tabular}
\end{center}
\end{table}
\subsection{Reaction path}

\begin{figure*}[!ht]
\centering
\includegraphics[scale=2.8]{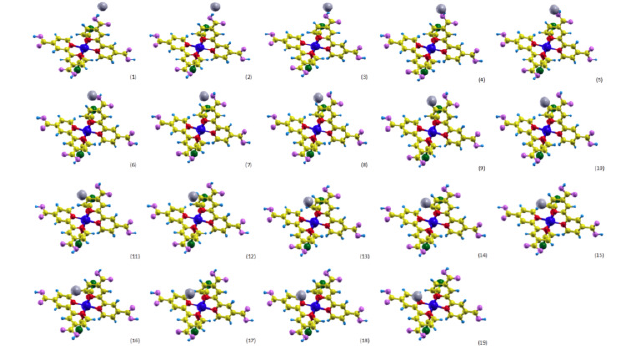}
\caption{
 configuration formed during the path of reaction \ref{equ:TS-1}. The first structure is 
 [$\rm N3^+ ... I_{Ru}^-$], the last structure is [$\rm N3^+ ... I_{OH}^-$], 
 and the 10th structure is TS-1.
}
\label{nebpic}
\end{figure*}

As previously mentioned, the dye regeneration mechanism happens through the formation of 
intermediate complexes and transition states. Since determination of lowest energy path is the 
one of the most important issue which we are dealing with, it can be useful to caculate it with 
Nudged Elastic Band (NEB) method. 
By using NEB method, it is possible to determine Minimum 
Energy Path (MEP), saddle point energy (maximum potential energy along the MEP), activation 
energy barrier, estimated transition rate and also configuration of atoms during the transition. 
Minimum Energy Path is calculated for reaction which states in eq.~\ref{equ:TS-1} 
by considering 19 images. In this 
method, a path with the lowest energy is determined for the reaction then a string of images of 
the system are created. In fact, images of the system are the intermediate configurations of 
system. Finally, the images are connected together through the hypothetical springs with the 
same spring constant \cite{Mills1998}, \cite{HenkelmanG.JohannessonG.andJonsson2002}. The 
representation of reaction path, from the reactant configuration to the 
product configuration, is formed. $\rm [N3^+...I_{Ru}^-]$ and $\rm [N3^+...I_{OH}^-]$ complexes are 
considered as the initial and final images, respectively and remain constant during the run. 
Also, intermediate images are obtained by using interpolation. It results an energy for each 
images. Minimum Energy Path, obtained from NEB method, is displayed in Fig. \ref{nebpath}.

\begin{align}
&\rm [N3^+...I_{Ru}^-] \hskip 0.3cm \stackrel{TS-1}\longrightarrow \hskip 0.3cm [N3^+...I_{OH}^-]
\label{equ:TS-1}
\end{align}

It can be seen from Fig. \ref{nebpath} that $\rm [N3^+...I_{Ru}^-]$ complex has a higher energy 
level than $\rm [N3^+...I_{OH}^-]$ complex, so $\rm [N3^+...I_{OH}^-]$ complex is more stable than 
$\rm [N3^+...I_{Ru}^-]$ complex and the occurrence of reaction is possible. Also, the second image 
has higher energy than its adjacent images. This increase in energy indicating weak bond formed 
during the reaction path, which is not unexpected. Moreover, tenth image has a maximum energy 
among other images and we introduce it as the transition state of reaction. The activation energy 
obtained by this method is presented in Table \ref{tab:activationE} and is compared with the 
activation energy obtained from other calculations. By comparing the values presented in 
Table \ref{tab:activationE}, we can see that the activation energy obtained from Nudged Elastic 
Band method has a better accordance with the activation energy reported in other studies, hence 
it can be said that the NEB method is a good and reliable method to determine the reaction path.
\\
Since in another reactions, the two reactants are converted to one product, calculation of the 
activation energy is not possible by using NEB method. In this case there are infinite situations 
for the location of the two reactants towards each other which is impossible to consider all 
situations, hence NEB method was used just for a reaction in which TS-1 transition state is 
formed.

As previously mentioned, one of the advantages of NEB method is determination of the 
configuration of atoms in any part of the reaction path. Due to the number of images that we 
have considered, 19 configuration is determined using interpolation. These configurations are 
displayed in Fig. \ref{nebpic}. As we know, TS-1 structure is a transition state that is 
achieved from conversion of [$\rm N3^+ ... I_{Ru}^-$] to [$\rm N3^+ ... I_{OH}^-$]. Also it can be seen 
clearly from Fig. \ref{nebpic} that iodine has moved from its place in [$\rm N3^+ ... I_{Ru}^-$] 
to that in [$\rm N3^+ ... I_{OH}^-$].

\section{CONCLUSION}

Density functional calculations were performed to investigate the structural, electronic, 
vibrational and optical properties of $\rm N3$ dye molecule, using both Quantum Espresso and 
FHI-aims Computational packages. We found that complexes of $\rm N3$ and transition states have 
magnetic properties. The obtained results from calculated IR spectrum indicate that $\rm N3$ dye 
molecule is a stable structure. Also, we have chosen the most stable of $\rm [dye^+...I_-]$ 
complexes for our calculations. We calculated optical gap of $\rm N3$ dye molecule using 
optical spectrum, then compared with experimental gap. The conformity of optical gap and 
experimental gap demonstrate the success of Liouville-Lanczos approach in calculating the optical gap and 
determination of optical gap. Also, we determine the minimum energy path of reaction 
\ref{equ:TS-1} and its activation energy, using Nudged Elastic Band method. It was discussed 
that the tenth image is $\rm TS-1$. In summery, investigation of details about mechanism of 
interactions between $\rm I/I_3^-$ and $\rm N3$ dye molecule can help us to a better understanding of dye 
regeneration and provide factors for raising efficiency or in the other words, it could be paving 
the way to achieve higher efficiency.

\section{ACKNOWLEDGMENTS}

We would like to acknowledge the Isfahan University of Technology. The authors gratefully 
acknowledge the Sheikh Bahaei National High Performance Computing Center (SBNHPCC) for 
providing computing facilities and time. SBNHPCC is supported by scientific and technological 
department of presidential office and Isfahan University of Technology (IUT).

\appendix*
\section{Photovoltaic mechanism}

The stable complexes formed through the following reactions, 
depends on the status of iodine in the complexes.

\begin{enumerate}

\item
Reactions involving the formation of complexes $\rm [N3^+...I_{Ru}^-]$ and $\rm N3I_{Ru}-I^-$:
 \begin{align}
&\rm N3+h\nu \longrightarrow N3^*,
\\
&\rm N3^* \longrightarrow N3^++e^-,
\\
& \rm I_3^-+2e^- \longrightarrow 3I^-,
\\
&\rm N3^++I^- \longrightarrow [N3^+...I_{Ru}^-],
\\
& \rm [N3^+...I_{Ru}^-]+I^- \longrightarrow N3I_{Ru}-I^-,
\\
& \rm N3I_{Ru}-I^- \longrightarrow N3+I_2^-,
\\
& 2I_2^- \longrightarrow I^-+I_3^-,
\end{align}
\item
Reactions involving the formation of complexes $\rm [N3^+...I_{OH}^-]$ and $\rm N3I_{OH}-I^-$:
\begin{align}
&\rm  N3+h\nu \longrightarrow N3^*,
\\
&\rm N3^* \longrightarrow N3^++e^-,
\\
& \rm I_3^-+2e^- \longrightarrow 3I^-,
\\
& \rm N3^++I^- \longrightarrow [N3^+...I_{OH}^-],
\\
& \rm[N3^+...I_{OH}^-]+I^- \longrightarrow N3I_{OH}-I^-,
\\
&\rm  N3I_{OH}-I^- \longrightarrow N3+I_2^-,
\\
&\rm 2I_2^- \longrightarrow I^-+I_3^-,
\end{align}
\end{enumerate}

\begin{figure}[!ht]
\centering
\includegraphics[scale=1.2]{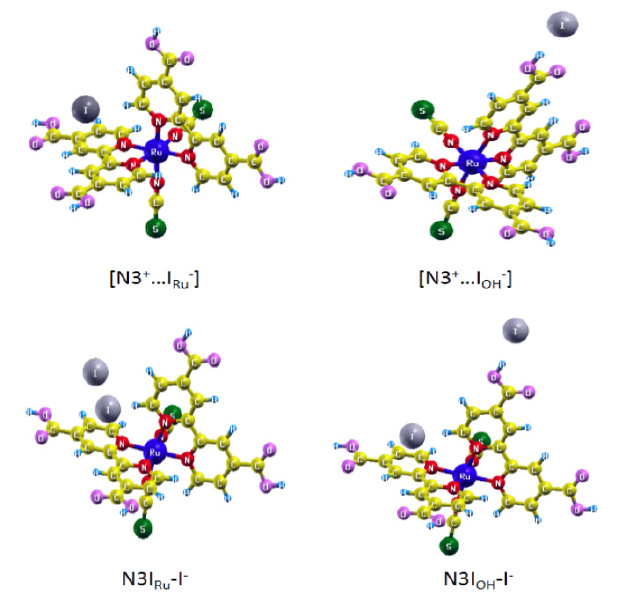}
\caption{
Optimized structure of $\rm N3^+I^-$ ([$\rm N3^+ ... I_{Ru}^-$], [$\rm N3^+ ... I_{OH}^-$]) 
and $\rm N3I_2^-$ ($\rm N3I_{Ru}-I^-$, $\rm N3I_{OH}-I^-$) complexes.
}
\label{complex}
\end{figure}

The optimized structures of $\rm N3^+I^-$ ([$\rm N3^+ ... I_{Ru}^-$], [$\rm N3^+ ... I_{OH}^-$]) 
and $\rm N3I_2^-$ ($\rm N3I_{Ru}-I^-$, $\rm N3I_{OH}-I^-$) complexes are shown in Fig. \ref{complex}.

During this reactions, the five transition states will be formed which are as follows:
\begin{align}
& \rm [N3^+...I_{Ru}^-] \hskip 0.3cm \stackrel{TS-1}\longrightarrow \hskip 0.3cm [N3^+...I_{OH}^-],
\\
& \rm [N3^+...I_{Ru}^-]+I^- \hskip 0.3cm \stackrel{TS-2}\longrightarrow \hskip 0.3cm N3I_{Ru}-I^-,
\\
& \rm [N3^+...I_{OH}^-]+I^- \hskip 0.3cm \stackrel{TS-3}\longrightarrow \hskip 0.3cm N3I_{OH}-I^-,
\\
& \rm N3I_{Ru}-I^- \hskip 0.3cm \stackrel{TS-4}\longrightarrow \hskip 0.3cm N3+I_2^-,
\\
& \rm N3I_{OH}-I^- \hskip 0.3cm \stackrel{TS-5}\longrightarrow \hskip 0.3cm N3+I_2^-,
\end{align}

\bibliography{N3}

\begin{thebibliography}{24}%
\makeatletter
\providecommand \@ifxundefined [1]{%
 \@ifx{#1\undefined}
}%
\providecommand \@ifnum [1]{%
 \ifnum #1\expandafter \@firstoftwo
 \else \expandafter \@secondoftwo
 \fi
}%
\providecommand \@ifx [1]{%
 \ifx #1\expandafter \@firstoftwo
 \else \expandafter \@secondoftwo
 \fi
}%
\providecommand \natexlab [1]{#1}%
\providecommand \enquote  [1]{``#1''}%
\providecommand \bibnamefont  [1]{#1}%
\providecommand \bibfnamefont [1]{#1}%
\providecommand \citenamefont [1]{#1}%
\providecommand \href@noop [0]{\@secondoftwo}%
\providecommand \href [0]{\begingroup \@sanitize@url \@href}%
\providecommand \@href[1]{\@@startlink{#1}\@@href}%
\providecommand \@@href[1]{\endgroup#1\@@endlink}%
\providecommand \@sanitize@url [0]{\catcode `\\12\catcode `\$12\catcode
  `\&12\catcode `\#12\catcode `\^12\catcode `\_12\catcode `\%12\relax}%
\providecommand \@@startlink[1]{}%
\providecommand \@@endlink[0]{}%
\providecommand \url  [0]{\begingroup\@sanitize@url \@url }%
\providecommand \@url [1]{\endgroup\@href {#1}{\urlprefix }}%
\providecommand \urlprefix  [0]{URL }%
\providecommand \Eprint [0]{\href }%
\providecommand \doibase [0]{http://dx.doi.org/}%
\providecommand \selectlanguage [0]{\@gobble}%
\providecommand \bibinfo  [0]{\@secondoftwo}%
\providecommand \bibfield  [0]{\@secondoftwo}%
\providecommand \translation [1]{[#1]}%
\providecommand \BibitemOpen [0]{}%
\providecommand \bibitemStop [0]{}%
\providecommand \bibitemNoStop [0]{.\EOS\space}%
\providecommand \EOS [0]{\spacefactor3000\relax}%
\providecommand \BibitemShut  [1]{\csname bibitem#1\endcsname}%
\let\auto@bib@innerbib\@empty
\bibitem [{\citenamefont {Panwar}\ \emph {et~al.}(2011)\citenamefont {Panwar},
  \citenamefont {Kaushik},\ and\ \citenamefont {Kothari}}]{panwar2011role}%
  \BibitemOpen
  \bibfield  {author} {\bibinfo {author} {\bibfnamefont {N.}~\bibnamefont
  {Panwar}}, \bibinfo {author} {\bibfnamefont {S.}~\bibnamefont {Kaushik}}, \
  and\ \bibinfo {author} {\bibfnamefont {S.}~\bibnamefont {Kothari}},\
  }\href@noop {} {\bibfield  {journal} {\bibinfo  {journal} {Renewable and
  Sustainable Energy Reviews}\ }\textbf {\bibinfo {volume} {15}},\ \bibinfo
  {pages} {1513} (\bibinfo {year} {2011})}\BibitemShut {NoStop}%
\bibitem [{\citenamefont {Bagher}\ \emph {et~al.}(2015)\citenamefont {Bagher},
  \citenamefont {Mahmoud}, \citenamefont {Vahid},\ and\ \citenamefont
  {Mohsen}}]{Bagher2015}%
  \BibitemOpen
  \bibfield  {author} {\bibinfo {author} {\bibfnamefont {A.~M.}\ \bibnamefont
  {Bagher}}, \bibinfo {author} {\bibfnamefont {M.}~\bibnamefont {Mahmoud}},
  \bibinfo {author} {\bibfnamefont {A.}~\bibnamefont {Vahid}}, \ and\ \bibinfo
  {author} {\bibfnamefont {M.}~\bibnamefont {Mohsen}},\ }\href {\doibase
  10.11648/j.ajop.20150305.17} {\bibfield  {journal} {\bibinfo  {journal}
  {American Journal of Optics and Photonics}\ }\textbf {\bibinfo {volume}
  {3(5)}},\ \bibinfo {pages} {94} (\bibinfo {year} {2015})}\BibitemShut
  {NoStop}%
\bibitem [{\citenamefont {Green}\ and\ \citenamefont
  {Emery}(2015)}]{Green2014}%
  \BibitemOpen
  \bibfield  {author} {\bibinfo {author} {\bibfnamefont {M.}~\bibnamefont
  {Green}}\ and\ \bibinfo {author} {\bibfnamefont {K.}~\bibnamefont {Emery}},\
  }\href {\doibase 10.1002/pip} {\bibfield  {journal} {\bibinfo  {journal}
  {Progress in \ldots}\ }\textbf {\bibinfo {volume} {22}},\ \bibinfo {pages}
  {701} (\bibinfo {year} {2015})}\BibitemShut {NoStop}%
\bibitem [{\citenamefont {Crabtree}\ and\ \citenamefont
  {Lewis}(2007)}]{crabtree2007solar}%
  \BibitemOpen
  \bibfield  {author} {\bibinfo {author} {\bibfnamefont {G.~W.}\ \bibnamefont
  {Crabtree}}\ and\ \bibinfo {author} {\bibfnamefont {N.~S.}\ \bibnamefont
  {Lewis}},\ }\href@noop {} {\bibfield  {journal} {\bibinfo  {journal} {Physics
  today}\ }\textbf {\bibinfo {volume} {60}},\ \bibinfo {pages} {37} (\bibinfo
  {year} {2007})}\BibitemShut {NoStop}%
\bibitem [{\citenamefont {Gr{\"a}tzel}(2006)}]{gratzel2006photovoltaic}%
  \BibitemOpen
  \bibfield  {author} {\bibinfo {author} {\bibfnamefont {M.}~\bibnamefont
  {Gr{\"a}tzel}},\ }\href@noop {} {\bibfield  {journal} {\bibinfo  {journal}
  {Comptes Rendus Chimie}\ }\textbf {\bibinfo {volume} {9}},\ \bibinfo {pages}
  {578} (\bibinfo {year} {2006})}\BibitemShut {NoStop}%
\bibitem [{\citenamefont {Pastore}\ \emph {et~al.}(2012)\citenamefont
  {Pastore}, \citenamefont {Mosconi},\ and\ \citenamefont
  {De~Angelis}}]{pastore2012computational}%
  \BibitemOpen
  \bibfield  {author} {\bibinfo {author} {\bibfnamefont {M.}~\bibnamefont
  {Pastore}}, \bibinfo {author} {\bibfnamefont {E.}~\bibnamefont {Mosconi}}, \
  and\ \bibinfo {author} {\bibfnamefont {F.}~\bibnamefont {De~Angelis}},\
  }\href@noop {} {\bibfield  {journal} {\bibinfo  {journal} {The Journal of
  Physical Chemistry C}\ }\textbf {\bibinfo {volume} {116}},\ \bibinfo {pages}
  {5965} (\bibinfo {year} {2012})}\BibitemShut {NoStop}%
\bibitem [{\citenamefont {{Gr\"{a}tzel, M}}(2003)}]{Gratzel2003}%
  \BibitemOpen
  \bibfield  {author} {\bibinfo {author} {\bibnamefont {{Gr\"{a}tzel, M}}},\
  }\href@noop {} {\bibfield  {journal} {\bibinfo  {journal} {Journal of
  Photochemistry and Photobiology C: Photochemistry Reviews}\ }\textbf
  {\bibinfo {volume} {4}},\ \bibinfo {pages} {145} (\bibinfo {year}
  {2003})}\BibitemShut {NoStop}%
\bibitem [{\citenamefont {Pizzoli}\ \emph {et~al.}(2012)\citenamefont
  {Pizzoli}, \citenamefont {Lobello}, \citenamefont {Carlotti}, \citenamefont
  {Elisei}, \citenamefont {Nazeeruddin}, \citenamefont {Vitillaro},\ and\
  \citenamefont {De~Angelis}}]{pizzoli2012acid}%
  \BibitemOpen
  \bibfield  {author} {\bibinfo {author} {\bibfnamefont {G.}~\bibnamefont
  {Pizzoli}}, \bibinfo {author} {\bibfnamefont {M.~G.}\ \bibnamefont
  {Lobello}}, \bibinfo {author} {\bibfnamefont {B.}~\bibnamefont {Carlotti}},
  \bibinfo {author} {\bibfnamefont {F.}~\bibnamefont {Elisei}}, \bibinfo
  {author} {\bibfnamefont {M.~K.}\ \bibnamefont {Nazeeruddin}}, \bibinfo
  {author} {\bibfnamefont {G.}~\bibnamefont {Vitillaro}}, \ and\ \bibinfo
  {author} {\bibfnamefont {F.}~\bibnamefont {De~Angelis}},\ }\href@noop {}
  {\bibfield  {journal} {\bibinfo  {journal} {Dalton Transactions}\ }\textbf
  {\bibinfo {volume} {41}},\ \bibinfo {pages} {11841} (\bibinfo {year}
  {2012})}\BibitemShut {NoStop}%
\bibitem [{\citenamefont {De~Angelis}\ \emph {et~al.}(2004)\citenamefont
  {De~Angelis}, \citenamefont {Fantacci},\ and\ \citenamefont
  {Selloni}}]{de2004time}%
  \BibitemOpen
  \bibfield  {author} {\bibinfo {author} {\bibfnamefont {F.}~\bibnamefont
  {De~Angelis}}, \bibinfo {author} {\bibfnamefont {S.}~\bibnamefont
  {Fantacci}}, \ and\ \bibinfo {author} {\bibfnamefont {A.}~\bibnamefont
  {Selloni}},\ }\href@noop {} {\bibfield  {journal} {\bibinfo  {journal}
  {Chemical physics letters}\ }\textbf {\bibinfo {volume} {389}},\ \bibinfo
  {pages} {204} (\bibinfo {year} {2004})}\BibitemShut {NoStop}%
\bibitem [{\citenamefont {Boschloo}\ and\ \citenamefont
  {Hagfeldt}(2009)}]{boschloo2009characteristics}%
  \BibitemOpen
  \bibfield  {author} {\bibinfo {author} {\bibfnamefont {G.}~\bibnamefont
  {Boschloo}}\ and\ \bibinfo {author} {\bibfnamefont {A.}~\bibnamefont
  {Hagfeldt}},\ }\href@noop {} {\bibfield  {journal} {\bibinfo  {journal}
  {Accounts of Chemical Research}\ }\textbf {\bibinfo {volume} {42}},\ \bibinfo
  {pages} {1819} (\bibinfo {year} {2009})}\BibitemShut {NoStop}%
\bibitem [{\citenamefont {Asaduzzaman}\ and\ \citenamefont
  {Schreckenbach}(2011)}]{Asaduzzaman2011}%
  \BibitemOpen
  \bibfield  {author} {\bibinfo {author} {\bibfnamefont {A.}~\bibnamefont
  {Asaduzzaman}}\ and\ \bibinfo {author} {\bibfnamefont {G.}~\bibnamefont
  {Schreckenbach}},\ }\href
  {http://pubs.rsc.org/en/content/articlehtml/2011/cp/c1cp21168d} {\bibfield
  {journal} {\bibinfo  {journal} {Physical Chemistry Chemical \ldots}\ }\textbf
  {\bibinfo {volume} {13}},\ \bibinfo {pages} {15148} (\bibinfo {year}
  {2011})}\BibitemShut {NoStop}%
\bibitem [{\citenamefont {Clifford}\ \emph {et~al.}(2007)\citenamefont
  {Clifford}, \citenamefont {Palomares}, \citenamefont {Nazeeruddin},
  \citenamefont {Gr{\"a}tzel},\ and\ \citenamefont
  {Durrant}}]{clifford2007dye}%
  \BibitemOpen
  \bibfield  {author} {\bibinfo {author} {\bibfnamefont {J.~N.}\ \bibnamefont
  {Clifford}}, \bibinfo {author} {\bibfnamefont {E.}~\bibnamefont {Palomares}},
  \bibinfo {author} {\bibfnamefont {M.~K.}\ \bibnamefont {Nazeeruddin}},
  \bibinfo {author} {\bibfnamefont {M.}~\bibnamefont {Gr{\"a}tzel}}, \ and\
  \bibinfo {author} {\bibfnamefont {J.~R.}\ \bibnamefont {Durrant}},\
  }\href@noop {} {\bibfield  {journal} {\bibinfo  {journal} {The Journal of
  Physical Chemistry C}\ }\textbf {\bibinfo {volume} {111}},\ \bibinfo {pages}
  {6561} (\bibinfo {year} {2007})}\BibitemShut {NoStop}%
\bibitem [{\citenamefont {Xie}\ \emph {et~al.}(2014)\citenamefont {Xie},
  \citenamefont {Chen}, \citenamefont {Bai}, \citenamefont {Wei},\ and\
  \citenamefont {Zhang}}]{xie2014theoretical}%
  \BibitemOpen
  \bibfield  {author} {\bibinfo {author} {\bibfnamefont {M.}~\bibnamefont
  {Xie}}, \bibinfo {author} {\bibfnamefont {J.}~\bibnamefont {Chen}}, \bibinfo
  {author} {\bibfnamefont {F.-Q.}\ \bibnamefont {Bai}}, \bibinfo {author}
  {\bibfnamefont {W.}~\bibnamefont {Wei}}, \ and\ \bibinfo {author}
  {\bibfnamefont {H.-X.}\ \bibnamefont {Zhang}},\ }\href@noop {} {\bibfield
  {journal} {\bibinfo  {journal} {The Journal of Physical Chemistry A}\
  }\textbf {\bibinfo {volume} {118}},\ \bibinfo {pages} {2244} (\bibinfo {year}
  {2014})}\BibitemShut {NoStop}%
\bibitem [{\citenamefont {Giannozzi}\ \emph {et~al.}(2009)\citenamefont
  {Giannozzi}, \citenamefont {Baroni}, \citenamefont {Bonini}, \citenamefont
  {Calandra}, \citenamefont {Car}, \citenamefont {Cavazzoni}, \citenamefont
  {Ceresoli}, \citenamefont {Chiarotti}, \citenamefont {Cococcioni},
  \citenamefont {Dabo}, \citenamefont {{Dal Corso}}, \citenamefont
  {de~Gironcoli}, \citenamefont {Fabris}, \citenamefont {Fratesi},
  \citenamefont {Gebauer}, \citenamefont {Gerstmann}, \citenamefont
  {Gougoussis}, \citenamefont {Kokalj}, \citenamefont {Lazzeri}, \citenamefont
  {Martin-Samos}, \citenamefont {Marzari}, \citenamefont {Mauri}, \citenamefont
  {Mazzarello}, \citenamefont {Paolini}, \citenamefont {Pasquarello},
  \citenamefont {Paulatto}, \citenamefont {Sbraccia}, \citenamefont {Scandolo},
  \citenamefont {Sclauzero}, \citenamefont {Seitsonen}, \citenamefont
  {Smogunov}, \citenamefont {Umari},\ and\ \citenamefont
  {Wentzcovitch}}]{Giannozzi2009}%
  \BibitemOpen
  \bibfield  {author} {\bibinfo {author} {\bibfnamefont {P.}~\bibnamefont
  {Giannozzi}}, \bibinfo {author} {\bibfnamefont {S.}~\bibnamefont {Baroni}},
  \bibinfo {author} {\bibfnamefont {N.}~\bibnamefont {Bonini}}, \bibinfo
  {author} {\bibfnamefont {M.}~\bibnamefont {Calandra}}, \bibinfo {author}
  {\bibfnamefont {R.}~\bibnamefont {Car}}, \bibinfo {author} {\bibfnamefont
  {C.}~\bibnamefont {Cavazzoni}}, \bibinfo {author} {\bibfnamefont
  {D.}~\bibnamefont {Ceresoli}}, \bibinfo {author} {\bibfnamefont {G.~L.}\
  \bibnamefont {Chiarotti}}, \bibinfo {author} {\bibfnamefont {M.}~\bibnamefont
  {Cococcioni}}, \bibinfo {author} {\bibfnamefont {I.}~\bibnamefont {Dabo}},
  \bibinfo {author} {\bibfnamefont {A.}~\bibnamefont {{Dal Corso}}}, \bibinfo
  {author} {\bibfnamefont {S.}~\bibnamefont {de~Gironcoli}}, \bibinfo {author}
  {\bibfnamefont {S.}~\bibnamefont {Fabris}}, \bibinfo {author} {\bibfnamefont
  {G.}~\bibnamefont {Fratesi}}, \bibinfo {author} {\bibfnamefont
  {R.}~\bibnamefont {Gebauer}}, \bibinfo {author} {\bibfnamefont
  {U.}~\bibnamefont {Gerstmann}}, \bibinfo {author} {\bibfnamefont
  {C.}~\bibnamefont {Gougoussis}}, \bibinfo {author} {\bibfnamefont
  {A.}~\bibnamefont {Kokalj}}, \bibinfo {author} {\bibfnamefont
  {M.}~\bibnamefont {Lazzeri}}, \bibinfo {author} {\bibfnamefont
  {L.}~\bibnamefont {Martin-Samos}}, \bibinfo {author} {\bibfnamefont
  {N.}~\bibnamefont {Marzari}}, \bibinfo {author} {\bibfnamefont
  {F.}~\bibnamefont {Mauri}}, \bibinfo {author} {\bibfnamefont
  {R.}~\bibnamefont {Mazzarello}}, \bibinfo {author} {\bibfnamefont
  {S.}~\bibnamefont {Paolini}}, \bibinfo {author} {\bibfnamefont
  {A.}~\bibnamefont {Pasquarello}}, \bibinfo {author} {\bibfnamefont
  {L.}~\bibnamefont {Paulatto}}, \bibinfo {author} {\bibfnamefont
  {C.}~\bibnamefont {Sbraccia}}, \bibinfo {author} {\bibfnamefont
  {S.}~\bibnamefont {Scandolo}}, \bibinfo {author} {\bibfnamefont
  {G.}~\bibnamefont {Sclauzero}}, \bibinfo {author} {\bibfnamefont {A.~P.}\
  \bibnamefont {Seitsonen}}, \bibinfo {author} {\bibfnamefont {A.}~\bibnamefont
  {Smogunov}}, \bibinfo {author} {\bibfnamefont {P.}~\bibnamefont {Umari}}, \
  and\ \bibinfo {author} {\bibfnamefont {R.~M.}\ \bibnamefont {Wentzcovitch}},\
  }\href {http://iopscience.iop.org/0953-8984/21/39/395502
  http://arxiv.org/abs/0906.2569 http://www.ncbi.nlm.nih.gov/pubmed/21832390}
  {\bibfield  {journal} {\bibinfo  {journal} {Journal of physics. Condensed
  matter}\ }\textbf {\bibinfo {volume} {21}},\ \bibinfo {pages} {395502}
  (\bibinfo {year} {2009})}\BibitemShut {NoStop}%
\bibitem [{\citenamefont {Blum}\ \emph {et~al.}(2009)\citenamefont {Blum},
  \citenamefont {Gehrke}, \citenamefont {Hanke},\ and\ \citenamefont
  {Havu}}]{Blum2009}%
  \BibitemOpen
  \bibfield  {author} {\bibinfo {author} {\bibfnamefont {V.}~\bibnamefont
  {Blum}}, \bibinfo {author} {\bibfnamefont {R.}~\bibnamefont {Gehrke}},
  \bibinfo {author} {\bibfnamefont {F.}~\bibnamefont {Hanke}}, \ and\ \bibinfo
  {author} {\bibfnamefont {P.}~\bibnamefont {Havu}},\ }\href
  {http://www.sciencedirect.com/science/article/pii/S0010465509002033}
  {\bibfield  {journal} {\bibinfo  {journal} {Computer Physics \ldots}\ ,\
  \bibinfo {pages} {1}} (\bibinfo {year} {2009})}\BibitemShut {NoStop}%
\bibitem [{\citenamefont {Perdew}\ \emph {et~al.}(1996)\citenamefont {Perdew},
  \citenamefont {Burke},\ and\ \citenamefont
  {Ernzerhof}}]{PhysRevLett.77.3865}%
  \BibitemOpen
  \bibfield  {author} {\bibinfo {author} {\bibfnamefont {J.~P.}\ \bibnamefont
  {Perdew}}, \bibinfo {author} {\bibfnamefont {K.}~\bibnamefont {Burke}}, \
  and\ \bibinfo {author} {\bibfnamefont {M.}~\bibnamefont {Ernzerhof}},\ }\href
  {\doibase 10.1103/PhysRevLett.77.3865} {\bibfield  {journal} {\bibinfo
  {journal} {Phys. Rev. Lett.}\ }\textbf {\bibinfo {volume} {77}},\ \bibinfo
  {pages} {3865} (\bibinfo {year} {1996})}\BibitemShut {NoStop}%
\bibitem [{\citenamefont {Malc{\i}o{\u{g}}lu}\ \emph
  {et~al.}(2011)\citenamefont {Malc{\i}o{\u{g}}lu}, \citenamefont {Gebauer},
  \citenamefont {Rocca},\ and\ \citenamefont
  {Baroni}}]{malciouglu2011turbotddft}%
  \BibitemOpen
  \bibfield  {author} {\bibinfo {author} {\bibfnamefont {O.~B.}\ \bibnamefont
  {Malc{\i}o{\u{g}}lu}}, \bibinfo {author} {\bibfnamefont {R.}~\bibnamefont
  {Gebauer}}, \bibinfo {author} {\bibfnamefont {D.}~\bibnamefont {Rocca}}, \
  and\ \bibinfo {author} {\bibfnamefont {S.}~\bibnamefont {Baroni}},\
  }\href@noop {} {\bibfield  {journal} {\bibinfo  {journal} {Computer Physics
  Communications}\ }\textbf {\bibinfo {volume} {182}},\ \bibinfo {pages} {1744}
  (\bibinfo {year} {2011})}\BibitemShut {NoStop}%
\bibitem [{\citenamefont {Ullrich}\ and\ \citenamefont
  {Yang}(2014)}]{ullrich2014brief}%
  \BibitemOpen
  \bibfield  {author} {\bibinfo {author} {\bibfnamefont {C.~A.}\ \bibnamefont
  {Ullrich}}\ and\ \bibinfo {author} {\bibfnamefont {Z.-h.}\ \bibnamefont
  {Yang}},\ }\href@noop {} {\bibfield  {journal} {\bibinfo  {journal}
  {Brazilian Journal of Physics}\ }\textbf {\bibinfo {volume} {44}},\ \bibinfo
  {pages} {154} (\bibinfo {year} {2014})}\BibitemShut {NoStop}%
\bibitem [{\citenamefont {Caspersen}\ and\ \citenamefont
  {Carter}(2005)}]{Caspersen2005}%
  \BibitemOpen
  \bibfield  {author} {\bibinfo {author} {\bibfnamefont {K.}~\bibnamefont
  {Caspersen}}\ and\ \bibinfo {author} {\bibfnamefont {E.}~\bibnamefont
  {Carter}},\ }\href {\doibase 10.1073/pnas.0408127102} {\bibfield  {journal}
  {\bibinfo  {journal} {Proceedings of the National Academy of Sciences of the
  United States of America}\ }\textbf {\bibinfo {volume} {102}},\ \bibinfo
  {pages} {6738} (\bibinfo {year} {2005})},\ \bibinfo {note} {cited By
  31}\BibitemShut {NoStop}%
\bibitem [{\citenamefont {Bersch}\ and\ \citenamefont
  {Rangan}(2008)}]{Bersch2008}%
  \BibitemOpen
  \bibfield  {author} {\bibinfo {author} {\bibfnamefont {E.}~\bibnamefont
  {Bersch}}\ and\ \bibinfo {author} {\bibfnamefont {S.}~\bibnamefont
  {Rangan}},\ }\href {http://adsabs.harvard.edu/abs/2008APS..MARW36006B}
  {\bibfield  {journal} {\bibinfo  {journal} {APS Meeting \ldots}\ }\textbf
  {\bibinfo {volume} {1}},\ \bibinfo {pages} {36006} (\bibinfo {year}
  {2008})}\BibitemShut {NoStop}%
\bibitem [{\citenamefont {Bai}\ \emph {et~al.}(2014)\citenamefont {Bai},
  \citenamefont {Mora-Ser\'{o}}, \citenamefont {{De Angelis}}, \citenamefont
  {Bisquert},\ and\ \citenamefont {Wang}}]{Bai2014}%
  \BibitemOpen
  \bibfield  {author} {\bibinfo {author} {\bibfnamefont {Y.}~\bibnamefont
  {Bai}}, \bibinfo {author} {\bibfnamefont {I.}~\bibnamefont {Mora-Ser\'{o}}},
  \bibinfo {author} {\bibfnamefont {F.}~\bibnamefont {{De Angelis}}}, \bibinfo
  {author} {\bibfnamefont {J.}~\bibnamefont {Bisquert}}, \ and\ \bibinfo
  {author} {\bibfnamefont {P.}~\bibnamefont {Wang}},\ }\href {\doibase
  10.1021/cr400606n} {\bibfield  {journal} {\bibinfo  {journal} {Chemical
  reviews}\ }\textbf {\bibinfo {volume} {114}},\ \bibinfo {pages} {10095}
  (\bibinfo {year} {2014})}\BibitemShut {NoStop}%
\bibitem [{\citenamefont {Schiffmann}\ \emph {et~al.}()\citenamefont
  {Schiffmann}, \citenamefont {VandeVondele}, \citenamefont {Hutter},
  \citenamefont {Wirz}, \citenamefont {Urakawa},\ and\ \citenamefont
  {Baiker}}]{Schiffmann2010}%
  \BibitemOpen
  \bibfield  {author} {\bibinfo {author} {\bibfnamefont {F.}~\bibnamefont
  {Schiffmann}}, \bibinfo {author} {\bibfnamefont {J.}~\bibnamefont
  {VandeVondele}}, \bibinfo {author} {\bibfnamefont {J.}~\bibnamefont
  {Hutter}}, \bibinfo {author} {\bibfnamefont {R.}~\bibnamefont {Wirz}},
  \bibinfo {author} {\bibfnamefont {A.}~\bibnamefont {Urakawa}}, \ and\
  \bibinfo {author} {\bibfnamefont {A.}~\bibnamefont {Baiker}},\ }\href@noop {}
  {\ }\BibitemShut {NoStop}%
\bibitem [{\citenamefont {Mills}\ and\ \citenamefont
  {Jacobsen}(1998)}]{Mills1998}%
  \BibitemOpen
  \bibfield  {author} {\bibinfo {author} {\bibfnamefont {G.}~\bibnamefont
  {Mills}}\ and\ \bibinfo {author} {\bibfnamefont {K.~W.}\ \bibnamefont
  {Jacobsen}},\ }in\ \href@noop {} {\emph {\bibinfo {booktitle} {In Classical
  and quantum dynamics in condensed phase simulations}}}\ (\bibinfo
  {publisher} {University of Washington},\ \bibinfo {year} {1998})\ pp.\
  \bibinfo {pages} {385--404}\BibitemShut {NoStop}%
\bibitem [{\citenamefont {{Henkelman, G., Johannesson, G. and
  Jonsson}}(2002)}]{HenkelmanG.JohannessonG.andJonsson2002}%
  \BibitemOpen
  \bibfield  {author} {\bibinfo {author} {\bibfnamefont {H.}~\bibnamefont
  {{Henkelman, G., Johannesson, G. and Jonsson}}},\ }in\ \href@noop {} {\emph
  {\bibinfo {booktitle} {in Theoretical Methods in Condensed Phase
  Chemistry}}}\ (\bibinfo {year} {2002})\ pp.\ \bibinfo {pages}
  {269--300}\BibitemShut {NoStop}%
\end{thebibliography}%
\end{document}